\documentclass[a4paper,12pt,oneside]{article}
\pdfoutput=1 
\usepackage[english]{babel}
\usepackage{amsmath}
\usepackage{amssymb}
\usepackage{epsf}
\usepackage{graphicx}
\usepackage{cite}

\usepackage[section,above,below]{placeins} 
\usepackage[unicode,colorlinks,citecolor=blue]{hyperref}

\usepackage{tocenter}
\FromMargins[f]{15mm}{15mm}{20mm}{20mm}

\renewcommand{\Im}{\mathop{\mathrm{Im}}\nolimits}

\DeclareMathOperator{\sign}{sgn}

\graphicspath{{img/}}

\title{Cherenkov and parametric (quasi-Cherenkov) radiation from relativistic charged particles moving in crystals formed by metallic wires}

\author{Baryshevsky V.G., Gurnevich E.A.\\
Research Institute for Nuclear Problems, Belarusian State
University,\\
11 Bobruiskaya Str., Minsk 220030, Belarus;\\
e-mail: bar$@$inp.bsu.by; genichgurn$@$gmail.com}

\date{}

\begin{document}
\maketitle

\begin{abstract}
Until recently, the interaction of electromagnetic waves with
crystals built from parallel metallic wires (wire media)  was analyzed in the
approximation of isotropic scattering of the electromagnetic wave by a single wire.
However, if the wires are thick ($kR\sim 1$), electromagnetic wave scattering by a wire
is anisotropic, i.e.,  the scattering amplitude depends on the scattering angle.
In this work, we  derive the equations that describe  diffraction of electromagnetic
waves and spontaneous emission of charged particles in wire media, and take into account the
angular dependence of scattering amplitude. Numerical solutions of these equations show that
the radiation intensity increases as the wire radius is increased and achieves
its maximal value in the range $kR\sim 1$. The case when the condition $kR\sim 1$
is fulfilled in the THz frequency range is considered in detail.
The calculations show that the instantaneous power of Cherenkov and parametric
(quasi-Cherenkov) radiations from electron bunches in the crystal 
can  be tens--hundreds megawatts, i.e., high enough to allow experimental observation
as well as possible practical applications. 

\end{abstract}

\section*{Introduction}

Emission of photons by relativistic charged particles moving in
natural or artificial spatially-periodic structures (photonic
crystals, metamaterials)  has come under intensive theoretical and
experimental investigation in recent years
\cite{PXR,Baryshevsky,Gurnevich2009,Gurnevich2010,Baryshevsky2015,Vorobev2012,shiffler2010cerenkov}.
It has been found that  spatial-periodic structure of crystals is
facilitating for new  mechanisms of radiation from a uniformly
moving particle to occur in addition to already known Cherenkov
and transition radiations: Smith-Purcell effect (diffraction radiation and
resonance radiation) \cite{smith_purcell,Bolotovsky,Ter-mikaelyan1960,Ter_mikaelyan1961}
and parametric X-ray radiation \cite{PXR}.

Parametric radiation in X-ray frequency range (PXR) caused by
particles moving in \textit{natural} crystals has been
theoretically and experimentally studied in numerous works (see
\cite{PXR}) and the references therein. Parametric radiation  from
a relativistic particle moving in  a {\it photonic} crystal  has
been considered in \cite{Baryshevsky,Baryshevsky2015}.
The theoretical description of PXR is based on Ewald's and von
Laue's dynamical theory of X-ray diffraction. It is important to
note that here the perturbation theory well applies to the
description of photon scattering by a single atom, whereas
in microwave and optical ranges the perturbation theory does not
always apply to describe photon scattering by scatterers that form
a photonic crystal (ball, wire, etc.).
Nevertheless, as it has been demonstrated
in \cite{Baryshevsky}, it is possible to derive
equations defining process of dynamical diffraction  in photonic crystals
and to describe
the emission of photons from relativistic particles moving in such crystals.

Radiation produced by charged particles moving in crystals built
from parallel metallic wires has been studied in
\cite{Baryshevsky,Baryshevsky2015,bar_molch2009,FEL2009grid,FirstGridExp1,Vorobev2012,radiationRev3,radiationRev4,radiationRev6,radiationRev7}.
The authors of
\cite{Vorobev2012,radiationRev3,radiationRev4,radiationRev6,radiationRev7}
have considered the case when the wavelength is much greater than
the crystal period, and so the diffraction conditions are not
fulfilled. Here  the crystal was presented as an equivalent
uniform medium  characterized by certain permittivity and
permeability tensors. Inversely, the authors of
\cite{Baryshevsky,Baryshevsky2015,bar_molch2009,FEL2009grid,FirstGridExp1}
have not confine themselves to the analysis of long-wave
approximation, because  diffraction in crystals is paramount for
the considered radiation mechanisms (parametric, diffraction).
According to the results reported in \cite{Baryshevsky2015}, when
the wavelength $\lambda=2\pi/k$ becomes comparable with the wire
radius $R$ ($kR \sim 1$) a noticeable increase in the intensity of
parametric radiation is observed. As a result,  say,  electron
bunches with  $n_e\sim 10^{9}$--$10^{11}$  that are produced  
through laser acceleration can generate GW-level THz pulses in
such crystals \cite{Baryshevsky2015}.

Let us note here that until now (see
\cite{Baryshevsky,Vorobev2012,radiationRev3,radiationRev4,radiationRev6},
a review article  \cite{BelovReview2012} and the reference
therein),  the authors  concerned themselves only with the case
when $kR\ll 1$, where scattering of the electromagnetic wave with
a ``parallel'' polarization (vector $\mathbf{E}$ is parallel to the
axes of the wires) by a single wire is isotropic, and scattering
of a wave with a ``perpendicular'' polarization can be neglected.
The analysis in \cite{Baryshevsky2015} relies on extrapolation of
the results obtained for the theory valid at $kR\ll 1$ to the
frequency range $kR\sim 1$, where, generally speaking, scattering
by a single wire is anisotropic (the scattering amplitude
depends on the scattering angle).

The equations describing the dynamical diffraction theory in 2D crystals that
are valid  for 
the case of anisotropic scatering by a single constituent element of the crystal (e.g. wire)
were first obtained
in \cite{Gurnevich2012}. In this paper, we use the theory
developed in \cite{Gurnevich2012} to give a detailed analysis of
refraction and diffraction of waves in crystals built from
metallic wires in the case when $kR\sim 1$. We derive the
equations  that describe spontaneous radiation from charged
particles moving in such crystals with due account of angular
dependence of the scattering amplitude. Numerical solution of the
derived equations shows that, as concluded in
\cite{Baryshevsky2015}, the radiation intensity increases with
increasing wire radius, achieving its maximum in the range $kR\sim
1$. We give a special consideration to the case when the condition
$kR\sim 1$ is fulfilled in THz range of frequencies.

The paper is arranged as follows. The first section describes the
general approach that we take to find the
characteristics of radiation produced by a charged particle moving
in arbitrary targets (photonic crystals).
The second section considers the theory of diffraction in photonic
crystals built from metallic wires and derives the dispersion
equation describing the possible types of waves in the
crystal that also  holds true in the case when the wires cannot be
regarded as thin ($kR\sim 1$).
The third section analyzes radiation produced by a charged particle
moving in crystals built from metallic wires, at different $kR$.

\section{Emission of photons by a charged particle moving in the crystal}
\label{sec:radiation_general}

\begin{figure}[htp]
\centering
 \center{\includegraphics[height=8cm]{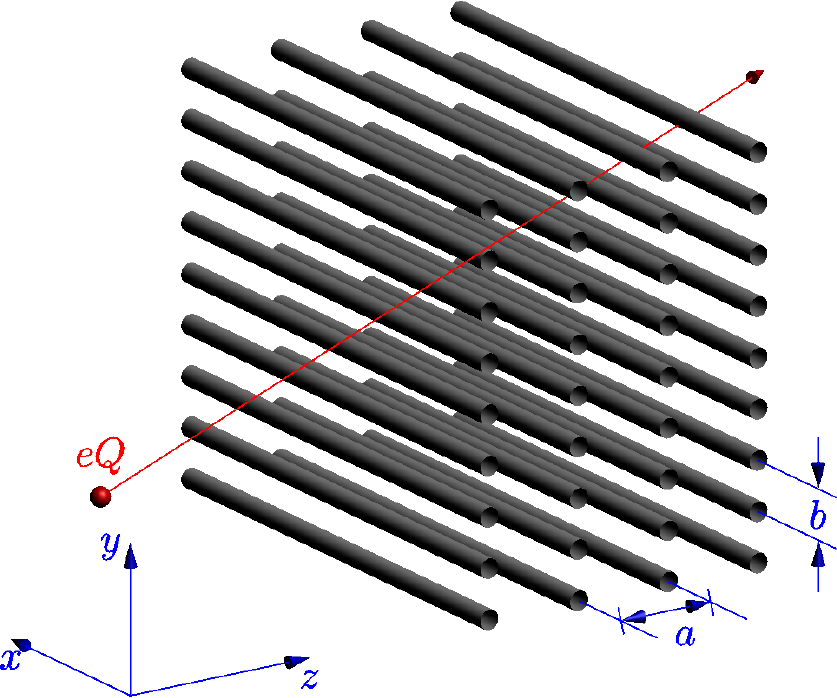}}
\caption{Photonic crystal built from metallic wires and the
used  coordinate system.}
\label{fig:crystal_geometry}
\end{figure}

Let a relativistic particle of charge $eQ$ move at a constant
velocity in a crystal built from parallel metallic wires, as shown
in Fig.~\ref{fig:crystal_geometry}. The crystal thickness $L$ is
assumed to be much less than its transverse dimensions and the
wire radius $R$ is much less than the crystal periods $a,b$.
Let us denote the  unit cell area by $\Omega_2=ab$.

To study the process of emission of electromagnetic waves  by a
charged particle, we shall use the general approach described in
\cite{PXR,baryshevsky2012high,Baryshevsky1997}. The spectral
density of radiation energy per unit solid angle
$W_{\mathbf{n}\omega}$ ($\mathbf{n}=\mathbf{k}/k$), differential
number of quanta  $\dfrac{d^2N}{d\omega d\Omega}\equiv
dN_{\mathbf{n}\omega}=W_{\mathbf{n}\omega}/\hbar\omega$, and the
polarization characteristics of radiation can be obtained readily
if we know the field $\mathbf{E}(\mathbf{r},\omega)$, produced by
a charged particle  at a large distance $\mathbf{r}$ from the target
(crystal):
\begin{equation}
 W_{\mathbf{n}\omega} = \frac{cr^2}{4\pi^2}\left|\mathbf{E}(\mathbf{r}, \omega) \right|^2,
\end{equation}
where  $c$ is the speed of light.

To find the field $\mathbf{E}(\mathbf{r},\omega)$, we need to
solve Maxwell's equations describing the interaction of particle
with a medium. The transverse solution can be found using the
Green function $G$ of these equations, which satisfies the
relationship (see \cite{PXR,baryshevsky2012high,Baryshevsky1997})
\begin{equation}
\label{berk_2.2} G=G_0+G_0\frac{\omega^2}{4\pi
c^2}(\hat{\varepsilon}-1)G.
\end{equation}
Here $G_0$ is the transverse Green function of Maxwell's equations
at $\hat{\varepsilon}=1$ (it is given, for example, in
\cite{Mors1953}) and $\hat{\varepsilon}$ is the permittivity
tensor of the medium. Using $G$, we can find the field we are
concerned with
\begin{equation}
\label{berk_2.3} E_n(\mathbf{r},\omega)=\int
G_{nl}(\mathbf{r},\mathbf{r}^{\prime},\omega)\frac{i\omega}{c^2}j_{0l}(\mathbf{r}^{\prime},\omega)d^3
r^{\prime},
\end{equation}
where $n,\, l =x,\, y,\, z$, and $j_{0l}(\mathbf{r},\omega)$ is the
Fourier transform of the $l$-th component of the current
produced by a moving charged particle.

According to   \cite{PXR, baryshevsky2012high},  the Green
function  at  $r\rightarrow \infty$ is expressed through the
solution of homogeneous Maxwell's equations containing at infinity
a  converging spherical wave $E^{(-)s}_{\mathbf{k}l}$:
\begin{equation}
\label{berk_2.4} \lim_{r\rightarrow \infty}
G_{nl}(\mathbf{r},\mathbf{r}^{\prime},\omega)=
\frac{e^{ikr}}{r}\sum\limits_s e^s_n E^{(-)s*}_{\mathbf{k}l}(\mathbf{r}^{\prime},\omega),\\
\end{equation}
where $\mathbf{k}= k\frac{\mathbf{r}}{r}$, $\mathbf{e}^{s}$ is
the unit polarization vector, $s=1,2$, and
$\mathbf{e}^{1}\perp\mathbf{e}^{2}\perp\mathbf{k}$. At
$r\rightarrow \infty$, the expression for wave
$E^{(-)s}_{\mathbf{k}l}(\mathbf{r}^{\prime})$ takes the form
\[
\mathbf{E}_k^{(-)s}(\mathbf{r}^{\prime}, \omega)=\mathbf{e}^{s}
e^{i\mathbf{k}\mathbf{r}^{\prime}}+\mbox{const}\frac{e^{-i {k}
{r}^{\prime}}}{r^{\prime}}.
\]
The solution $\mathbf{E}_{\mathbf{k}}^{(-)s} (\mathbf{r},\omega)$
 is associated with the
solution of homogeneous Maxwell's equations
$\mathbf{E}_{\mathbf{k}}^{(+)s}(\mathbf{r}, \omega)$
 that describes photon scattering
by a target and contains at infinity a
 diverging spherical wave
($\mathbf{E}_k^{(+)s}(\mathbf{r}, \omega)=\mathbf{e}^{\, \, s}
e^{i\mathbf{k}\mathbf{r}}+\mbox{const}\frac{e^{i {k} {r}}}{r}$),
\cite{PXR,baryshevsky2012high}:
\begin{equation}
\label{berk_2.5}
\mathbf{E}^{(-)s*}_{\mathbf{k}}=\mathbf{E}^{(+)s}_{-\mathbf{k}}.
\end{equation}
%

Using (\ref{berk_2.3}) and  (\ref{berk_2.4}), we obtain
\begin{equation}
\label{berk_2.6} E_n(\mathbf{r},
\omega)=\frac{e^{ikr}}{r}\frac{i\omega}{c^2} \sum\limits_s
e^{s}_n\int \mathbf{E}^{(-)s*}_{\mathbf{k}}(\mathbf{r}^{\prime},
\omega)\mathbf{j}_0(\mathbf{r}^{\prime}, \omega)d^3 r^{\prime}=
\frac{e^{ikr}}{r}\frac{i\omega}{c^2} \sum\limits_s
e^{s}_n\int \mathbf{E}^{(+)s}_{-\mathbf{k}}(\mathbf{r}^{\prime},
\omega)\mathbf{j}_0(\mathbf{r}^{\prime}, \omega)d^3 r^{\prime}.
\end{equation}
Then the spectral density of radiation for  photons with
polarization vector $\mathbf{e}^{s}$ can be written in the form:
\begin{equation}
\label{berk_2.7}
W_{\mathbf{n},\omega}^s=\frac{\omega^2}{4\pi^2c^3}\left|\int\mathbf{E}^{(-)s*}_{\mathbf{k}}(\mathbf{r},
\omega) \mathbf{j}_0(\mathbf{r}, \omega)d^3 r\right|^2,
\end{equation}
where
\begin{equation}
\label{berk_2.8} \mathbf{j}_0(\mathbf{r}, \omega)=\int e^{i\omega
t}\mathbf{j}_0(\mathbf{r}, t) dt=eQ \int e^{i\omega
t}\mathbf{v}(t)\delta(\mathbf{r}-\mathbf{r}(t))dt,
\end{equation}
and $\mathbf{v}(t)$ and $\mathbf{r}(t)$ are the particle velocity
and trajectory, respectively.

Substitution of (\ref{berk_2.8}) and \eqref{berk_2.5} into
(\ref{berk_2.7}) finally gives
\begin{equation}
\label{eq:Spec_ang_gen} dN^s_{\mathbf{n}, \omega}=\frac{e^2
Q^2\omega}{4\pi^2 \hbar c^3}\left|\int
\mathbf{E}^{(+)s}_{-\mathbf{k}}(\mathbf{r}(t),
\omega)\mathbf{v}(t) e^{i\omega t}dt\right|^2.
\end{equation}
 Integration in  \eqref{eq:Spec_ang_gen} is performed over the entire
domain of particle motion.

Thus, for the analysis of
radiation emitted by a
particle passing through a photonic crystal, we do not need
a complete solution of Maxwell's equations; suffice it
to know the solution of \textit{homogeneous} Maxwell's equations
describing plane-wave diffraction by the crystal.
Solving  homogeneous Maxwell's equations
 instead of inhomogeneous significantly simplifies the
analysis of the radiation problem and enables considering
different cases of radiation.

\section{Propagation of waves in a crystal built from parallel metallic wires}

Let us consider refraction and diffraction of electromagnetic
waves in a photonic crystal for the case when  $kR \sim 1$.
We shall start with solving the problem of plane-wave scattering
by a single wire (metallic cylinder), then consider scattering by
an one-dimensional grating (a single crystal plane) formed by
periodically spaced wires, and finally proceed to deriving the
dispersion equation for a infinite crystal.



\subsection{Amplitude of electromagnetic wave scattering by a wire}
Let a plane electromagnetic wave
$\mathbf{E}_0=\mathbf{e}_0e^{i\mathbf{k}_1\mathbf{r}}$
($\mathbf{e}_0$ is the polarization unit vector) be scattered by
an infinite cylinder of radius $R$. We shall assume that the
cylinder is placed in the medium whose permittivity and
permeability are
 $\varepsilon_1$ and $\mu_1$, respectively; we shall denote the permittivity and the permeability of the cylinder material
  by $\varepsilon_2$ and $\mu_2$, respectively. It is also assumed that the axis
of the cylinder is oriented along the $x$-axis of the Cartesian
rectangular coordinates and the wave vector of the incident wave
is $\mathbf{k}_1=(k_{1x},0,k_{1z})$. We shall also introduce a
polar coordinate system $(\rho, \varphi)$ in the $(z,y)$-plane
using the relations $z = \rho\cos\varphi$ and $y =
\rho\sin\varphi$.

It is necessary to consider two possible polarizations of the
incident  wave :
\begin{itemize}
    \item transverse electric (TE) polarization, when the vector $\mathbf{E}_0$ of the electric field strength
    is perpendicular to the axis of the cylinder  ($E_{0x} = 0$).  Hereinafter the quantities referring to this polarization will
    bear the index  ``$\perp$'';

    \item transverse magnetic (TM) polarization, when the vector $\mathbf{H}_0$ of the magnetic-field strength
    is perpendicular to  the axis of the cylinder   ($H_{0x} = 0$).
    Hereinafter the quantities referring to this polarization will
    bear the index  ``$\parallel$'',
    since in this case vector $\mathbf{E}$ has a nonzero
    component which is parallel to the cylinder axis $x$.
\end{itemize}
If the incident wave is TM-polarized, then the $x$-component of
the field $\mathbf{E}$ can be presented as a series in terms of
cylindrical functions \cite{Nikolsky,BelRev25,Wait1955}:
\begin{equation}
E_x = \left\{
\begin{aligned}
\sum\limits_{n=-\infty}^{\infty} i^n J_n(k_{\rho}
\rho)e^{-in\varphi} +
\sum\limits_{n=-\infty}^{\infty} i^n a_n H_n(k_{\rho} \rho)e^{-in\varphi},\;\; \rho \geq R \\
\sum\limits_{n=-\infty}^{\infty} i^n b_n J_n(k'_{\rho}
\rho)e^{-in\varphi},\;\; \rho \leq R,
\end{aligned}
\right. \label{eq:Ex_decomp}
\end{equation}
where $J_n$ and  $H_n$ are the  Bessel cylindrical function of the
$n$-th order and the  Hankel cylindrical function of the first
kind of the $n$-th order, respectively,
$\mathbf{k}_{\rho}=(k_{1y}, k_{1z})$, $k_{\rho} =
\sqrt{k^2\varepsilon_1\mu_1-k_{1x}^2}$, $k'_{\rho} =
\sqrt{k^2\varepsilon_2\mu_2-k_{1x}^2}$, $k=\omega/c$. For brevity,
we shall omit the factors $e^{ik_{1x}x}$ in
expressions for the field \eqref{eq:Ex_decomp}, as well as
hereinafter in this paper.   The component $H_x$ of the magnetic
field can be represented in a similar way:
\begin{equation}
H_x = \left\{
\begin{aligned}
\sum\limits_{n=-\infty}^{\infty} i^n c_n H_n(k_{\rho} \rho)e^{-in\varphi},\;\; \rho \geq R \\
\sum\limits_{n=-\infty}^{\infty} i^n d_n J_n(k'_{\rho}
\rho)e^{-in\varphi},\;\; \rho \leq R.
\end{aligned}
\right. \label{eq:Hx_decomp}
\end{equation}
For  ÒÅ-polarization, the expansion \eqref{eq:Hx_decomp} must be
used for  $E_x$ and the expansion  \eqref{eq:Ex_decomp} for $H_x$.
Other components of  the fields are expressed in terms of
$E_x$ and  $H_x$ as follows:
\begin{equation}
\left\{
\begin{aligned}
\mathbf{E}_\perp & \equiv & (E_y,E_z) & = &
\frac{ik_{1x}}{k_\rho^{2}}\vec{\nabla}_\perp E_x -
\frac{ik\mu_1}{k_\rho^{2}}\mathbf{e}_x\times
\vec{\nabla}_\perp H_x \\
\mathbf{H}_\perp & \equiv &  (H_y,H_z) & = &
\frac{ik_{1x}}{k_\rho^{2}}\vec{\nabla}_\perp H_x +
\frac{ik\varepsilon_1}{k_\rho^{2}}\mathbf{e}_x\times
\vec{\nabla}_\perp E_x,
\end{aligned}
\right. \label{eq:field_perp_comps}
\end{equation}
where $\vec{\nabla}_\perp = \mathbf{e}_y\frac{\partial}{\partial
y} + \mathbf{e}_z\frac{\partial}{\partial z}$, and $(\mathbf{e}_x,
\mathbf{e}_y, \mathbf{e}_z)$ are the unit vectors of the
corresponding axes. Equations \eqref{eq:field_perp_comps} are
valid for  $\rho \geq R$; in the case when  $\rho < R$, we should
replace  $k_\rho \rightarrow k_\rho^{\prime}$,
$\varepsilon_1\rightarrow \varepsilon_2$, $\mu_1\rightarrow
\mu_2$. The unknown coefficients $a_n$, $b_n$, $c_n$, and $d_n$
are found from boundary conditions at $\rho=R$.

Let the condition  $|k_{1x}|\ll k_\rho$ be fulfilled. Then for the
coefficients $a_n^\parallel$ and $a_n^\perp$ (i.e., the expansion
coefficients for the electric field outside the cylinder in the
case of TM-polarization ($\parallel$) and for the magnetic field
in the case of TE-polarization ($\perp$)), we can obtain the
following approximate expressions (see~\cite{Nikolsky,Wait1955}):
\begin{equation}
 \left\{
 \begin{aligned}
a_n^\parallel \approx  \frac{-J_n(k'_\rho R)J'_n(k_\rho R) +
\sqrt{\frac{\varepsilon}{\mu}} J'_n(k'_\rho R)J_n(k_\rho
R)}{J_n(k'_\rho R)H'_n(k_\rho R) -
\sqrt{\frac{\varepsilon}{\mu}}
J'_n(k'_\rho R)H_n(k_\rho R)},\\
a_n^\perp \approx \frac{-J_n(k'_\rho R)J'_n(k_\rho R) +
\sqrt{\frac{\mu}{\varepsilon}} J'_n(k'_\rho R)J_n(k_\rho
R)}{J_n(k'_\rho R)H'_n(k_\rho R) -
\sqrt{\frac{\mu}{\varepsilon}} J'_n(k'_\rho R)H_n(k_\rho R)},
 \end{aligned}
 \right.
 \label{eq:an_coeffs}
\end{equation}
where $\varepsilon=\varepsilon_2/\varepsilon_1$,
$\mu=\mu_2/\mu_1$. Formulas  \eqref{eq:an_coeffs} are exact for
$n=0$. If $n\neq 0$, these formulas are exact only when
$k_{1x}=0$, whereas at small values of $k_{1x}$ provide the
relative error of the order of $\frac{|k_{1x}|}{k_\rho}$.

For a wire made from a nonmagnetic metal and  placed in a vacuum,
the permeability $\mu$ and the permittivity $\varepsilon$ in
\eqref{eq:an_coeffs} should be taken as $\mu=1$ and
$\varepsilon=1+i\frac{4\pi \sigma}{\omega}$, where $\sigma$ is the
conductivity of the metal. Let us write the expressions for the
coefficients $a_n$ in the case of perfectly conducting wires for
their particular simple form. Since  the permittivity
$\varepsilon\rightarrow i\infty$ as $\sigma \rightarrow \infty$,
then  from \eqref{eq:an_coeffs} we find

\begin{equation}
\left\{
\begin{aligned}
a_n^\parallel = -\frac{J_n(k_\rho R)}{H_n(k_\rho R)},\\
a_n^\perp = -\frac{J'_n(k_\rho R)}{H'_n(k_\rho R)}.
\end{aligned}
\right. \label{eq:an_coeffs_perfect}
\end{equation}

Let us pay attention to the fact that if the incident plane wave
is TM(TE)-polarized, then in the case of perfectly conducting
wire, the scattering field is also completely TM(TE)-polarized
($H_x = 0$ or $E_x=0$, respectively). But this is not so in the
 general case when   $\varepsilon$ is an arbitrary value. For
 example, when a TM-polarized wave is incident onto the wire (i.e.,
 $H_{x,inc}=0$), then in the series expansion of the scattering
 field $H_{x,sc}=\sum\limits_{n=-\infty}^{\infty}i^nc_n^\parallel
H_n(k_\rho \rho)e^{-in\varphi}$, the coefficients
$c_n^\parallel$, generally speaking, differ from zero (except for
$c_0^\parallel$).
Thus, in this example, the  TM-polarized wave incident  on the
wire  produces a diffracted one that  contains the components with
TM- and TE-polarizations with  their amplitudes determined by the
coefficients $a_n^\parallel$ and $c_n^\parallel$, respectively.
 It can be shown, however, that in the case under
 consideration, at $|k_{1x}|\ll k_\rho$, the amplitudes $|c_n|\ll
 |a_n|$, and so can be neglected.

\begin{figure}[h]
    \centering
    \includegraphics{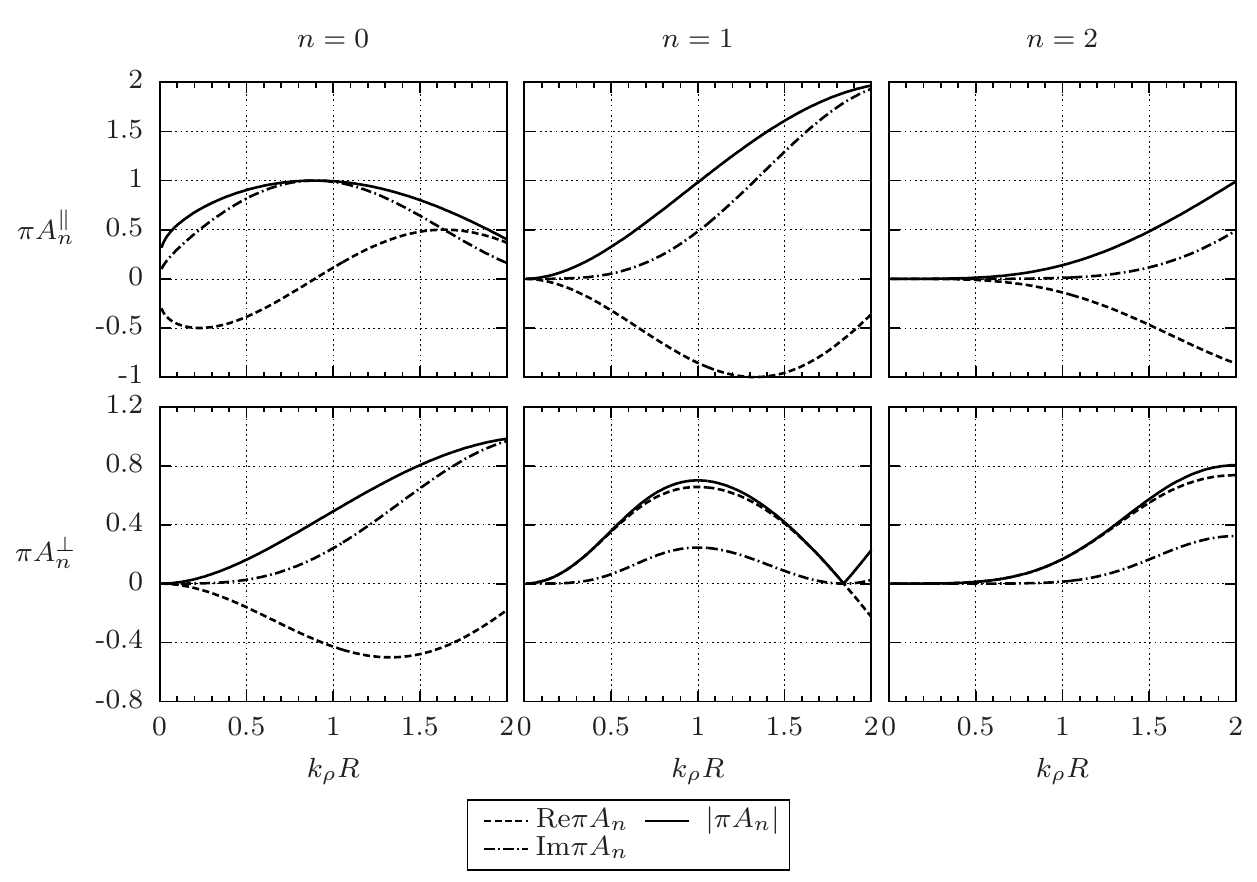}
    \caption{The amplitudes  $A_0$, $A_1$ and $A_2$ of electromagnetic wave
    scattering by a perfectly conducting wire versus $k_\rho R$
    for two polarizations}
    \label{fig:amplitudes}
\end{figure}

Using the asymptotic forms for Hankel functions of large argument
and the integral representation of Hankel functions
\cite{Mors1953,Janke}, the expressions for $E_x$ and $H_x$ at
large distances from the axis of the cylinder  ($k\rho \gg 1$) can
be written as
\begin{equation}
\Psi = \left\{
\begin{aligned}
E_x\\
H_x
\end{aligned}
\right\}= e^{i\mathbf{k}_{\rho}\pmb{\rho}} + \left\{
\begin{aligned}
A^\parallel(\varphi)\\
A^\perp(\varphi)
\end{aligned}
\right\}\times
\int\limits_{-\infty}^{\infty}\frac{e^{ik\sqrt{\rho^2+x^2}}}{\sqrt{\rho^2+x^2}}
\mathrm{d}x, \label{eq3:asymptotic}
\end{equation}
where
\begin{equation}
A^{\parallel (\perp)}(\varphi) =
-\frac{i}{\pi}\left(a_0^{\parallel (\perp)} +
2\sum\limits_{n=1}^{\infty}a_n^{\parallel
(\perp)}\cos(n\varphi)\right) \equiv \sum\limits_{n=0}^\infty
A_n^{\parallel(\perp)}\cos (n\varphi).
\label{eq:amplitude_expansion}
\end{equation}
Other unknown components of the fields can be expressed in terms
of  $E_x$ and $H_x$ using \eqref{eq:field_perp_comps}. Following
the analogy of a three-dimensional case, by $A(\varphi)$ we shall
mean the amplitude of scattering of the electromagnetic wave by a
cylinder at an angle $\varphi$ \cite{landau77quant}.

Figure~\ref{fig:amplitudes} exemplifies the expansion coefficients
$A_n$ for the  amplitudes  of scattering by a perfectly conducting
cylinder as a function of  $k_\rho R$ (they have a similar form
for a cylinder made from a finite-conductivity metal). It may be
seen that in the range $0<k_\rho R\lesssim 1$, we can take account
only of the expansion terms $n=0,1$ to consider scattering by a
cylinder, because other  terms are small. So the
 expression for  a wave scattered by a wire with the coordinates
 $\mathbf{\rho}_0=(y_0,z_0)$ can be  written in the form:
\begin{equation}
\Psi = e^{i\mathbf{k}_{\rho}\pmb{\rho}} + i\pi A_0 H_0(k_\rho
|\pmb{\rho}-\pmb{\rho}_0|) - \pi A_1 H_1(k_\rho
|\pmb{\rho}-\pmb{\rho}_0|) \cos (\mathbf{k}_\perp,
\pmb{\rho}-\pmb{\rho}_0), \label{eq:scattered_field_one}
\end{equation}
where the upper index of the scattering amplitude is omitted for
simplicity. In the case when  $k_\rho R \gtrsim 1$, we need to
consider many terms ($n=0,1,2,...$) in expansion of the scattering   
amplitude \eqref{eq:amplitude_expansion}, and the expression for a
wave scattered by the wire becomes quite complicated.  In our
further analysis we shall confine ourselves to the case when
$0<k_\rho R\lesssim 1$, and scattering by a single wire can be
described by equation \eqref{eq:scattered_field_one}.

\subsection{Scattering of electromagnetic waves by one-dimensional grating}

\begin{figure}[htp]
    \centering
    \includegraphics[height=5cm]{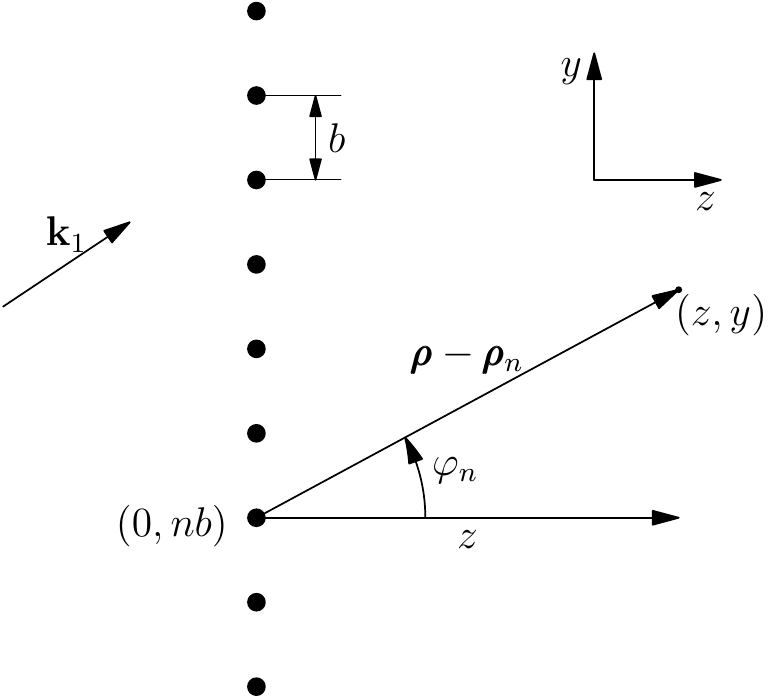}
    \caption{Scattering of a plane wave by a one-dimensional grating formed by metallic wires}
    \label{fig:one_dim_grating}
\end{figure}

Now, let us consider scattering of a plane wave by one-dimensional
grating formed by periodically spaced parallel wires
(Fig.~\ref{fig:one_dim_grating}).
Let the coordinates of the wire axes be $\pmb{\rho}_n=(z_n,y_n)$,
$y_n=bn$, $z_n=0$, where $b$ is the grating period, and the wave
vector of incident wave $\mathbf{k}_1=(k_x,k_y,k_z)$ (the index
``$1$'' on the components of the vector is omitted). The general
solution for the wave scattered by the grating has the form
\begin{multline}
\Psi = e^{i\mathbf{k}_{\rho}\pmb{\rho}} + i\pi
F_0\sum\limits_{n=-\infty}^{\infty}e^{ik_ybn}H_0(k_\rho
|\pmb{\rho}-\pmb{\rho}_n|)
-\pi F_1 \sum\limits_{n=-\infty}^{\infty}e^{ik_ybn}H_1(k_\rho |\pmb{\rho}-\pmb{\rho}_n|)\cos \varphi_n\\
-\pi F'_1 \sum\limits_{n=-\infty}^{\infty}e^{ik_ybn}H_1(k_\rho
|\pmb{\rho}-\pmb{\rho}_n|)\sin \varphi_n,
\label{eq:field_one_dim_grid_base}
\end{multline}
where summation is made over the coordinates $\pmb{\rho}_n$ of all
wires, and the amplitudes    $F_0$, $F_1$, and $F'_1$
are independent of the position of the wire (index $n$)
because the grating is assumed to be infinite.

The cylindrical wave of  amplitude $F(\varphi_n) =
e^{ik_ybn}(F_0+F_1\cos\varphi_n+F'_1\sin\varphi_n)$ diverging from
the $n$-th wire is produced through scattering by this wire of two
waves: 1)~the incident plane wave and 2)~the sum of cylindrical
waves (of amplitudes $F(\varphi_m)$, $m\neq n$) coming from all
other wires. Scattering of the initial plane wave by the wire is
described by the expression  \eqref{eq:scattered_field_one}; to
describe scattering of cylindrical waves, we can also use
\eqref{eq:scattered_field_one} provided that each cylindrical wave
is presented as a superposition of plane waves
\cite{Gurnevich2012}. In a similar manner as in
\cite{Gurnevich2012}, formulation a system of linear algebraic
equations for  $F_0$, $F_1$, and $F'_1$, enables us to find
\begin{equation}
\left\{
\begin{aligned}
F_0 = F_0(k_y, k_\rho) & = A_0\frac{1 + A_1(S_3+S_2k_y/k_\rho -S_1)}{1 - S_1A_0 - (S_1-S_3)A_1 - A_0A_1(S_1S_3+S_2^2 - S_1^2)},\\
F_1 = F_1(k_y, k_\rho) & = A_1\frac{k_z/k_\rho}{1 - A_1S_3},\\
F'_1 = F'_1(k_y, k_\rho) & = A_1\frac{k_y/k_\rho + A_0(S_2-S_1k_y/k_\rho)}{1 -
S_1A_0 - (S_1-S_3)A_1 - A_0A_1(S_1S_3+S_2^2 - S_1^2)},
\end{aligned}
\right. \label{eq:amplitudes_F}
\end{equation}
where  $S_1$, $S_2$, and $S_3$ are
\begin{multline}
S_1 = i\pi \sum\limits_{n=1}^{\infty} H_0(k_\rho bn)(e^{ik_ybn} +
e^{-ik_y bn}) = \frac{2i\pi}{k_{z0}b} - i\pi +2\left(\log
\frac{k_\rho b}{4\pi} + C\right) + \sum\limits_{n\neq 0}\left\{
\frac{2i\pi }{k_{zn}b} - \frac{1}{|n|} \right\}, \label{eq:S1_sum}
\end{multline}
\begin{equation}
S_2 = \pi \sum\limits_{n=1}^{\infty} H_1(k_\rho bn)(e^{ik_ybn} -
e^{-ik_y bn}) = -2\frac{k_y}{k_\rho} + \frac{i\pi}{k_\rho b}
\sum\limits_{n= -\infty}^{\infty} \left\{\frac{k_{y,-n}}{k_{z,-n}}
+ \frac{k_{yn}}{k_{zn}} \right\}, \label{eq:S2_sum}
\end{equation}
\begin{multline}
S_3 = i\pi \sum\limits_{n=1}^{\infty} \frac{H_1(k_\rho
bn)}{k_\rho b n}(e^{ik_ybn} + e^{-ik_y bn}) =
\frac{2i\pi}{k_\rho b}\frac{k_{z0}}{k_\rho} - \frac{i\pi}{2}
+\left(\log \frac{k_\rho b}{4\pi} + C - \frac{1}{2} +
\frac{k_y^2}{k_\rho^2} + \frac{2\pi^2}{3k_\rho^2 b^2}\right)
+\\+ \frac{2\pi}{k_\rho b}\sum\limits_{n\neq 0}\left\{
\frac{ik_{zn}}{k_\rho} + \frac{|k_{yn}|}{k_\rho} - \frac{k_\rho
b}{4\pi |n|} \right\}. \label{eq:S3_sum}
\end{multline}
In these formulas $C\approx 0.5772$ is the Euler constant, $k_{yn}
= k_y - 2\pi n/b$, and $k_{zn}=\sqrt{k_\rho^2 - k_{yn}^2}$. The
root is taken arithmetically, while in the case when the radicand
is negative, we assume that $\sqrt{-|\{...\}|} =
+i\sqrt{|\{...\}|}$. Let us note that  for $k_y b = \pi n$ the
expressions \eqref{eq:amplitudes_F} simplify appreciably, because
in this case $S_2=0$.

Let us concentrate on the analysis
of scattering of a TM-polarized wave.
Using the Poisson
summation formula in \eqref{eq:field_one_dim_grid_base}, we can
obtain the following expressions for the electric field
$\mathbf{E}$ in the case of scattering by a one-dimensional
grating  of a plane TM-polarized wave of unit amplitude:
\begin{equation}
\begin{aligned}
E_x & =  \frac{k_\rho}{k_1}\left\{e^{i\mathbf{k}_{\rho}\pmb{\rho}} +
\sum\limits_{n=-\infty}^{\infty} \left(F_0 +
F_1\frac{k_{zn}}{k_\rho}\sign z +
F'_1\frac{k_{yn}}{k_\rho} \right)
\frac{2i\pi}{k_{zn}b}e^{ik_{yn}y}e^{ik_{zn}|z|}\right\},\\
E_y & =
-\frac{k_x}{k_1}\left\{\frac{k_y}{k_\rho}e^{i\mathbf{k}_{\rho}\pmb{\rho}} +
\sum\limits_{n=-\infty}^{\infty} \left(F_0 +
F_1\frac{k_{zn}}{k_\rho}\sign z +
F'_1\frac{k_{yn}}{k_\rho} \right)
\frac{2i\pi}{k_{zn}b}\frac{k_{yn}}{k_\rho}e^{ik_{yn}y}e^{ik_{zn}|z|}\right\},\\
E_z & =  -\frac{k_x}{k_1}\left\{\frac{k_z}{k_\rho}e^{i\mathbf{k}_{\rho}\pmb{\rho}} +
\sum\limits_{n=-\infty}^{\infty} \left(F_0 +
F_1\frac{k_{zn}}{k_\rho}\sign z +
F'_1\frac{k_{yn}}{k_\rho} \right)
\frac{2i\pi}{k_{zn}b}\frac{k_{zn}}{k_\rho}
e^{ik_{yn}y}e^{ik_{zn}|z|}\sign z\right\}.
\end{aligned}
\label{eq:fields_one_dim_grid}
\end{equation}

To help grasp the physical meaning of the expressions
\eqref{eq:fields_one_dim_grid} more readily, we can write them in
a simpler, more compact form
\begin{equation}
\mathbf{E}(\mathbf{r}) = \mathbf{e}_0 e^{i\mathbf{k}_1\mathbf{r}}  + \sum\limits_{n=-\infty}^{\infty}
\mathbf{e}_n^{\pm} \Phi_n^{\pm} e^{i\mathbf{k}_n^{\pm}\mathbf{r}},
\label{eq:fields_one_dim_grid_vector}
\end{equation}
where $\mathbf{e}_0$ is the polarization unit vector of the wave
incident onto the grating,  $\mathbf{e}_n^{\pm} =
\frac{k_\rho}{k_1}\mathbf{e}_x - \frac{k_xk_{yn}}{k_1k_\rho}
\mathbf{e}_y \mp \frac{k_xk_{zn}}{k_1k_\rho}\mathbf{e}_z$ are the
polarization  unit vectors of waves diverging from the grating
(scattered waves), $\mathbf{k}_n^{\pm}=(k_x, k_{yn}, \pm k_{zn})$  are
their wave vectors, and  $\Phi_n^{\pm}$ are their amplitudes:
\begin{equation}
\left\{
\begin{aligned}
\Phi_n^{+} &= \frac{2i\pi}{k_{zn}b}\left(F_0 + F_1 \frac{k_{zn}}{k_\rho}
+ F'_1\frac{k_{yn}}{k_\rho}\right) = \frac{2i\pi}{k_{zn}b}F(\phi_n),\\
\Phi_n^{-} &= \frac{2i\pi}{k_{zn}b}\left(F_0 - F_1 \frac{k_{zn}}{k_\rho}
+ F'_1\frac{k_{yn}}{k_\rho}\right) = \frac{2i\pi}{k_{zn}b}F(\pi-\phi_n),
\end{aligned}
\right.
\label{eq:one_dim_grid_Phinew}
\end{equation}
where $F(\varphi)=F_0 + F_1\cos\varphi + F'_1\sin\varphi$ is the
effective amplitude of wave scattering by the wire of the grating,
$\sin\phi_n = \frac{k_{yn}}{k_\rho}$, $\cos\phi_n =
\frac{k_{zn}}{k_\rho}$. The sign ``$+$'' in
\eqref{eq:fields_one_dim_grid_vector}-\eqref{eq:one_dim_grid_Phinew}
refers to the case $z>0$, whereas the sign  ``$-$'' refers to the
case  $z<0$.

Formula \eqref{eq:fields_one_dim_grid_vector} reflects a simple
physical fact: in a general case,  diffraction of a plane wave
having a unit amplitude at a 1D periodic grating gives rise to a
set of diverging from the grating plane waves whose amplitudes are
$\Phi_n^{\pm}$. As follows from the expression for
$\mathbf{k}_n^{\pm}$, the $y$-components of wave vectors of  the
scattered waves differ from one another by the reciprocal
lattice vector  $\tau_y=\frac{2\pi n}{b}$. We shall note here that
for chaotically placed in the $z$-plane wires  (by contrast to
periodically placed) there would appear two plane waves;
transmitted through   and mirror-reflected from the wire array
(with wave vectors $\mathbf{k}_0^+\equiv\mathbf{k}_1=(k_x, k_y,
k_z)$ and $\mathbf{k}_0^-=(k_x, k_y, -k_z)$, respectively).

We should also state that at  $|k_{yn}|>k_\rho$, the wave with
index  $n$ in  \eqref{eq:fields_one_dim_grid_vector} is evanescent
wave, so at large distances from the grating (at sufficiently
 large $|z|$), summation in
\eqref{eq:fields_one_dim_grid_vector} should be confined only to
such values of $n$ at which  $|k_{yn}|<k_\rho$. However, in
analyzing the propagation of waves in a 2D crystal, it may be
essential that the evanescent waves be taken into account,
particularly if the crystal period along the $z$ direction is not
large enough, and so here we use the complete expressions for the
field \eqref{eq:fields_one_dim_grid}-\eqref{eq:fields_one_dim_grid_vector}.


\subsection{Diffraction and refraction in crystals at arbitrary scattering amplitudes}
\label{sec:diffr_and_refr} 
\begin{figure}[htp]
    \centering
    \includegraphics[width=0.7\linewidth]{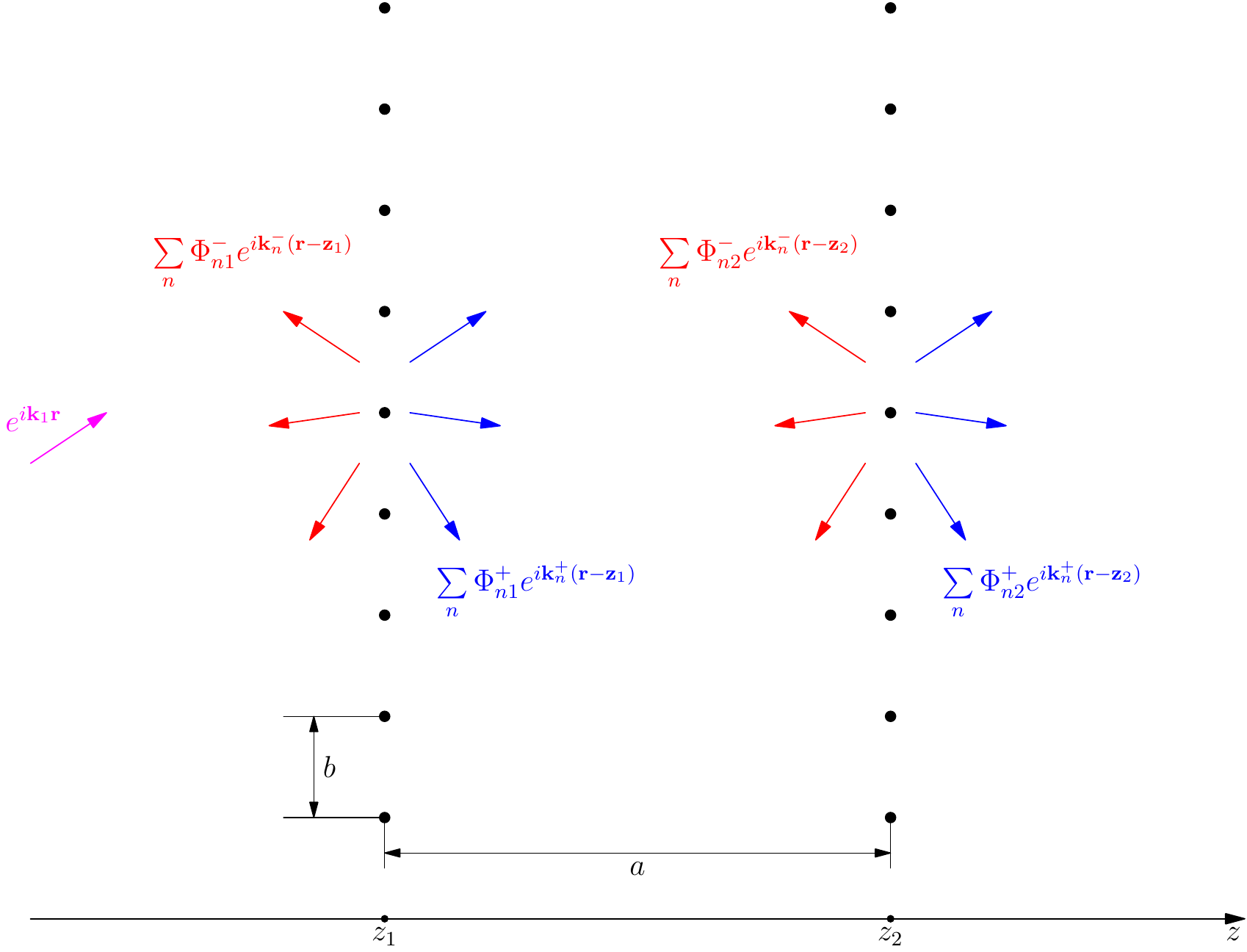}
    \caption{Scattering of a plane wave by two one-dimensional gratings}
    \label{fig:two_gratings}
\end{figure}

Let now a plane wave with TM
polarization  be scattered by two gratings instead of one. The
gratings are placed at a distance $a$ from one another in a medium
whose permittivity and permeability are   $\varepsilon_1$ and
$\mu_1$, respectively (Fig.~\ref{fig:two_gratings}). The general
solution has a similar form as
\eqref{eq:fields_one_dim_grid}-\eqref{eq:fields_one_dim_grid_vector}
i.e., for the $E_x$-component of the field $\mathbf{E}$ we can
write
\begin{equation}
E_x = \frac{k_\rho}{k_1}e^{i\mathbf{k}_1\mathbf{r}}
 + \frac{k_\rho}{k_1}\sum\limits_{n} \Phi_{n1}^\pm e^{i\mathbf{k}_n^\pm (\mathbf{r}-\mathbf{z}_1)}
 + \frac{k_\rho}{k_1}\sum\limits_{n} \Phi_{n2}^\pm e^{i\mathbf{k}_n^\pm (\mathbf{r}-\mathbf{z}_2)},
\end{equation}
where $\mathbf{z}_1= \mathbf{0}$, $\mathbf{z}_2=a\mathbf{e}_z$ are
the vectors defining the positions  of the first and second
gratings, respectively; the sign ``$+$'' or
``$-$'' in the first sum is chosen according to the sign of the
$z$-coordinate, whereas in the second sum the sign is chosen
according to the sign of the difference $(z-a)$. The remaining
components of the field can be expressed in terms of $E_x$ using
\eqref{eq:field_perp_comps} in a similar manner as was done
earlier in this paper.

The unknown amplitudes  $\Phi_{n1}^\pm$ and $\Phi_{n2}^\pm$ can be
expressed in terms of the amplitudes  $\Phi_n^\pm$ calculated for
the case of scattering by a single grating. We should bear in
mind here that each grating scatters not only  the initial plane
wave
$E_{x0}\equiv\Psi_0=\frac{k_\rho}{k_1}e^{i\mathbf{k}_1\mathbf{r}}$,
but also waves coming from the other grating  (see
Fig.~\ref{fig:two_gratings}). For example, the first grating
scatters the wave  $\Psi_0 + \Psi_2$, where
\begin{equation*}
\Psi_2 = \frac{k_\rho}{k_1}\sum\limits_{n} \Phi_{n2}^- e^{i\mathbf{k}_n^- (\mathbf{r}-\mathbf{z}_2)}.
\end{equation*}
As a result, we have a set of plane waves diverging from the first
grating:
\begin{equation*}
\Psi_1^{sc} = \frac{k_\rho}{k_1}\sum\limits_{n} \Phi_{n1}^\pm e^{i\mathbf{k}_n^\pm (\mathbf{r}-\mathbf{z}_1)}.
\end{equation*}
in a similar way, the second grating is affected by the wave
 $\Psi_0+\Psi_1$, where
\begin{equation*}
\Psi_1 = \frac{k_\rho}{k_1}\sum\limits_{n} \Phi_{n1}^+ e^{i\mathbf{k}_n^+ (\mathbf{r}-\mathbf{z}_1)},
\end{equation*}
and as a result, we have a scattered wave
\begin{equation*}
\Psi_2^{sc} = \frac{k_\rho}{k_1}\sum\limits_{n} \Phi_{n2}^\pm e^{i\mathbf{k}_n^\pm (\mathbf{r}-\mathbf{z}_2)}.
\end{equation*}
Knowing the law describing scattering of a unit-amplitude plane
wave by a single grating (see
 \eqref{eq:fields_one_dim_grid_vector}-\eqref{eq:one_dim_grid_Phinew}),
 we can write a system of linear algebraic equations for finding
  $\Phi_{n1}$ and  $\Phi_{n2}$.

Let us consider a very simple example of normal incidence of wave
onto a grating  when  $\mathbf{k}_1 = (0, 0, k_1)$ and
$k_\rho=k_1$. Let the wave number satisfies the condition
$k_1b<2\pi$; then in \eqref{eq:fields_one_dim_grid_vector} all the
waves with $|n|>0$ are evanescent and can be neglected if the distance
$a$ between the gratings is large enough. Taking into account that
for normal incidence  $F'_1=0$ (see \eqref{eq:amplitudes_F}) and
$\Phi_0^\pm=\frac{2i\pi}{k_1b}(F_0\pm F_1)$, expressions
\eqref{eq:fields_one_dim_grid_vector}--\eqref{eq:one_dim_grid_Phinew}
yield the following equation for scattering by a single grating
placed at  $z=z_0$
\begin{equation}
E_x \approx e^{ik_1z} + \frac{2i\pi}{k_1b}\left(F_0 + F_1\sign(z-z_0)\right)e^{ik_1|z-z_0|}.
\end{equation}
In the case of scattering by two gratings,
$\Phi_{0m}^\pm=\frac{2i\pi}{k_1b}(F_{0m}\pm F_{1m})$, where
$m=1,2$ and the earlier written formulas for  $\Psi_0$, $\Psi_1$,
$\Psi_2$ take the form
\begin{equation}
\Psi_0(z) = e^{ik_1 z},\;
\Psi_1(z) = \frac{2i\pi}{k_1b}\left(F_{01}+F_{11}\right)e^{ik_1|z-z_1|},\;
\Psi_2(z) = \frac{2i\pi}{k_1b}\left(F_{02}-F_{12}\right)e^{ik_1|z-z_2|}.
\end{equation}
Now we can write the sought system of equations as:
\begin{equation}
\left\{
\begin{aligned}
F_{01} & = F_0\cdot\left[\Psi_0(z_1) + \Psi_2(z_1)\right],\\
F_{11} & = F_1\cdot\left[\Psi_0(z_1) - \Psi_2(z_1)\right],\\
F_{02} & = F_0\cdot\left[\Psi_0(z_2) + \Psi_1(z_2)\right],\\
F_{12} & = F_1\cdot\left[\Psi_0(z_2) + \Psi_1(z_2)\right].
\end{aligned}
\right.
\label{eq:Phi_system_normal}
\end{equation}
Here the sign  ``$-$'' before the second term in the second
equation is because the wave $\Psi_2(z)$ is incident onto the
first grating from the positive values of $z$ (it propagate
in the direction opposite to that of the wave $\Psi_0$).

Taking account of evanescent waves and generalization of the obtained
results to the case of arbitrary (non-normal) incidence of the
initial wave onto the gratings requires cumbersome arithmetic
transformations but do not present  serious difficulties. The same
refers to the case when there are several ($M$) gratings instead
of two. The general solution for the field   $E_x$ in here has the   
form
\begin{equation}
\begin{aligned}
E_x & =  \frac{k_\rho}{k_1}\left\{e^{i\mathbf{k}_{\rho}\pmb{\rho}} +
\sum\limits_{n,m} \left(F_{0m} +
F_{1m}\frac{k_{zn}}{k_\rho}\sign (z-z_m) +
F'_{1m}\frac{k_{yn}}{k_\rho} \right)
\frac{2i\pi}{k_{zn}b} e^{ik_{yn}y}e^{ik_{zn}|z-z_m|}\right\}, 
\end{aligned}
\label{eq:Ex_field_crystal}
\end{equation}
where  $z_m=(m-1)a$, the summation over $m$ is made from 1 to $M$
(i.e., over all gratings), and  $F_{0m}$, $F_{1m}$, and $F'_{1m}$
are found from the system of linear equations:
\begin{equation}
\begin{aligned}
\frac{F_{0m}}{F_0} & = e^{ik_z z_m} + \sum\limits_{n\neq
m}\sum\limits_{l} e^{ik_{zl}|z_m-z_n|}
\frac{F_0(k_{yl})}{F_0(k_{y0})}
\frac{2i\pi }{k_{zl}b}\left(F_{0n} + F_{1n}\frac{k_{zl}}{k_\rho}\sign(m-n) + F'_{1n}\frac{k_{yl}}{k_\rho}\right),\\
\frac{F_{1m}}{F_1} & = e^{ik_z z_m} + \sum\limits_{n\neq
m}\sum\limits_{l} e^{ik_{zl}|z_m-z_n|}
\frac{F_1(k_{yl})}{F_1(k_{y0})}
\frac{2i\pi }{k_{zl}b}\left(F_{0n}\sign(m-n) + F_{1n}\frac{k_{zl}}{k_\rho} + F'_{1n}\frac{k_{yl}}{k_\rho}\sign(m-n)\right),\\
\frac{F'_{1m}}{F'_1} & = e^{ik_z z_m} + \sum\limits_{n\neq
m}\sum\limits_{l} e^{ik_{zl}|z_m-z_n|}
\frac{F'_1(k_{yl})}{F'_1(k_{y0})}
\frac{2i\pi }{k_{zl}b}\left(F_{0n} + F_{1n}\frac{k_{zl}}{k_\rho}\sign(m-n) + F'_{1n}\frac{k_{yl}}{k_\rho}\right).
\end{aligned}
\label{eq:Phi_system}
\end{equation}
This system of equations has a structure very similar to
\eqref{eq:Phi_system_normal} and considers scattering by each
($m$-th) grating of the initial plane wave $\Psi_0$  (the first
term on the right-hand side of \eqref{eq:Phi_system}) as well as
the waves coming from all other gratings (the second term). These
equations are more  awkward than \eqref{eq:Phi_system_normal}
because they consider all evanescent waves and the general case of
arbitrary (non-normal) incidence onto the grating.


Now let us proceed  from a set of $M$-number of one-dimensional
gratings to an infinite (in the $z$ direction) crystal. We shall
consider the problem  of finding the refractive index of an
infinite crystal as an eigenvalue problem as proposed by P.P.
Ewald in the development of dynamical theory of diffraction
 \cite{Pinsker,Ewald1962,James1948}. A coherent wave propagating
 through a crystal is a result of summation of elementary waves
 emitted by single wires. Production and propagation of these
 elementary  waves in an infinite, unbounded crystal should be
 considered as free oscillations (eigenmodes) of the
 system (crystal) rather than forced ones (excited by an external
 incident wave) \cite{Pinsker,Ewald1962,James1948}.
An important feature of this system is self-consistency manifested
in excitation of each wire 
 by the wave
field  formed by the superposition of elementary waves from all
other wires. The same is true not only for single wires, but also
for crystal planes (1D gratings discussed earlier in this
paper), i.e.,  each plane starts to emit waves under the effect of
the field induced by all other planes. Let us note that similar
reasoning  was used in the analysis of wave propagation in
crystals formed by anisotropically scattering centers in
\cite{Gurnevich2012,Gurnevich2013arxiv} and in \cite{Belov1} â for
the case of isotropic scattering.

The above can also be stated as follows. The solution of Maxwell's
equations  \eqref{eq:Ex_field_crystal} describes the field induced
in crystal  through scattering of a plane wave
$\mathbf{E}=\mathbf{e}_0 e^{i\mathbf{k}\mathbf{r}}$, and $F_{0m}$,
$F_{1m}$, and $F'_{1m}$ in  \eqref{eq:Ex_field_crystal} are the
solutions of the system of linear \textit{nonhomogeneous}
equations  \eqref{eq:Phi_system}, whose column (vector) of
constant terms is just determined by the amplitude of the incident
wave. To find the 
eigenmodes of 
infinite crystal,
we need to solve the appropriate system of \textit{homogenous}
equations   \cite{Agranovich1984}. According to Bloch's theorem,
we assume that $F_{0n}=F_{00}e^{iq_z z_n}$, $F_{1n}=F_{10}e^{iq_z
z_n}$, $F'_{1n}=F'_{10}e^{iq_z z_n}$, where $q_z$ is the unknown
$z$-component of the wave vector in the crystal, and $F_{00}$,
$F_{10}$, and  $F'_{10}$ are independent of $z_n$. Substitution of
these  expressions into  \eqref{eq:Phi_system} and elimination of
the constant terms $e^{ik_z z_m}$ corresponding to the wave
incident onto the crystal,
gives the system of linear equations  for  $F_{00}$, $F_{10}$, and
$F'_{10}$, having nonzero solution only when its determinant is
zero. After the summation over $n$ (it can be easily done by the
formula for geometric series), the substitution of
\eqref{eq:amplitudes_F} and some other transformations, we obtain
the dispersion equation for finding $q_z$ in the form
\begin{equation}
\det{D}=0, \label{eq:DE_full}
\end{equation}
where $D$ is  a certain $3\times 3$ matrix. Because in the general
case the elements of $D$ are quite cumbersome, we shall  define
them in the Appendix~\ref{appendix_A}.

By way of example, let us consider  the case of normal incidence
of the wave onto the crystal ($k_x=k_y=0$). The dispersion
equation  in this case takes the simplest form:
\begin{equation}
\frac{k_1b}{2\pi}=-\frac{C_1A_0}{1+i\pi A_0-S'_1 A_0} -
\frac{C_5A_1}{1+i\pi A_1/2-S'_3 A_1}
+\frac{2\pi}{k_1b}\frac{(C_3^2-C_1C_5)A_0A_1}{(1+i\pi A_0-S'_1
A_0)(1+i\pi A_1/2-S'_3 A_1)}, \label{eq:DU_normal}
\end{equation}
where $S'_1=\mathrm{Re}\, S_1$ and $S'_3=\mathrm{Re}\, S_3$,
whereas the functions  $C_1$, $C_3$, and $C_5$ depend on the
crystal periods $a$ and $b$ and the wave numbers
$k_1=k\sqrt{\varepsilon_1\mu_1}$ and $q$ (see
Appendix~\ref{appendix_A}). Let  $q=kn$, where  $n$ is the crystal
refractive index. If the condition  $|n^2-\varepsilon_1\mu_1|\ll
1$ is fulfilled, and $k_1a,k_1b < 2\pi$, then the functions $C_1$,
$C_3$, and $C_5$ will be approximately equal to
\begin{equation}
C_1\approx C_5 \approx -C_3 \approx -\frac{2\sqrt{\varepsilon_1\mu_1}}{ka(n^2-\varepsilon_1\mu_1)}.
\label{eq:C_simple}
\end{equation}
Substitution of these values into \eqref{eq:DU_normal} readily
gives the expression for the refractive index  $n$ and the
crystal's  effective dielectric  susceptibility $g_0$:
\begin{equation}
g_0 \equiv n^2 - 1 \approx \varepsilon_1\mu_1 - 1 + \frac{4\pi}{k^2 ab}\left\{ \frac{A_0}{1+i\pi
A_0-S'_1 A_0} + \frac{A_1}{1+i\pi A_1/2-S'_3 A_1}\right\},
\label{eq:g0_simp_v1}
\end{equation}
If $\varepsilon_1=\mu_1=1$ and the scattering amplitudes are small
(so as we can neglect  $S'_1A_0$ and  $S'_3A_1$), this expression,
in fact, coincides with the results obtained in
\cite{Gurnevich2012}. 

In a similar way we can solve the dispersion equation in the case
when the diffraction conditions in the crystal are fulfilled. Thus
derived expressions for effective polarizabilities $g_{\pmb{\tau}}$
(coefficients of Fourier expansion of the crystal's  effective
dielectric  susceptibility in terms reciprocal lattice vectors
$\pmb{\tau}$) are also the same as those given in
\cite{Gurnevich2012} (see Appendix~\ref{appendix_A}).

If the condition  $|n^2-\varepsilon_1\mu_1|\ll 1$ for the
refractive index is not fulfilled and if the scattering amplitude
is not small ($|\pi A_{0,1}|\lesssim 1$), we need to use
\eqref{eq:SC_sums_full} in Appendix~\ref{appendix_A} instead of the
approximate expressions \eqref{eq:C_simple}.  Of course in this
case, the solutions of the dispersion equations  (complete
\eqref{eq:DE_full} or simplified \eqref{eq:DU_normal}, depending
on the geometry of the problem)  shall be found numerically.

\begin{figure}[htp]
    \centering
    \includegraphics[width=0.47\linewidth]{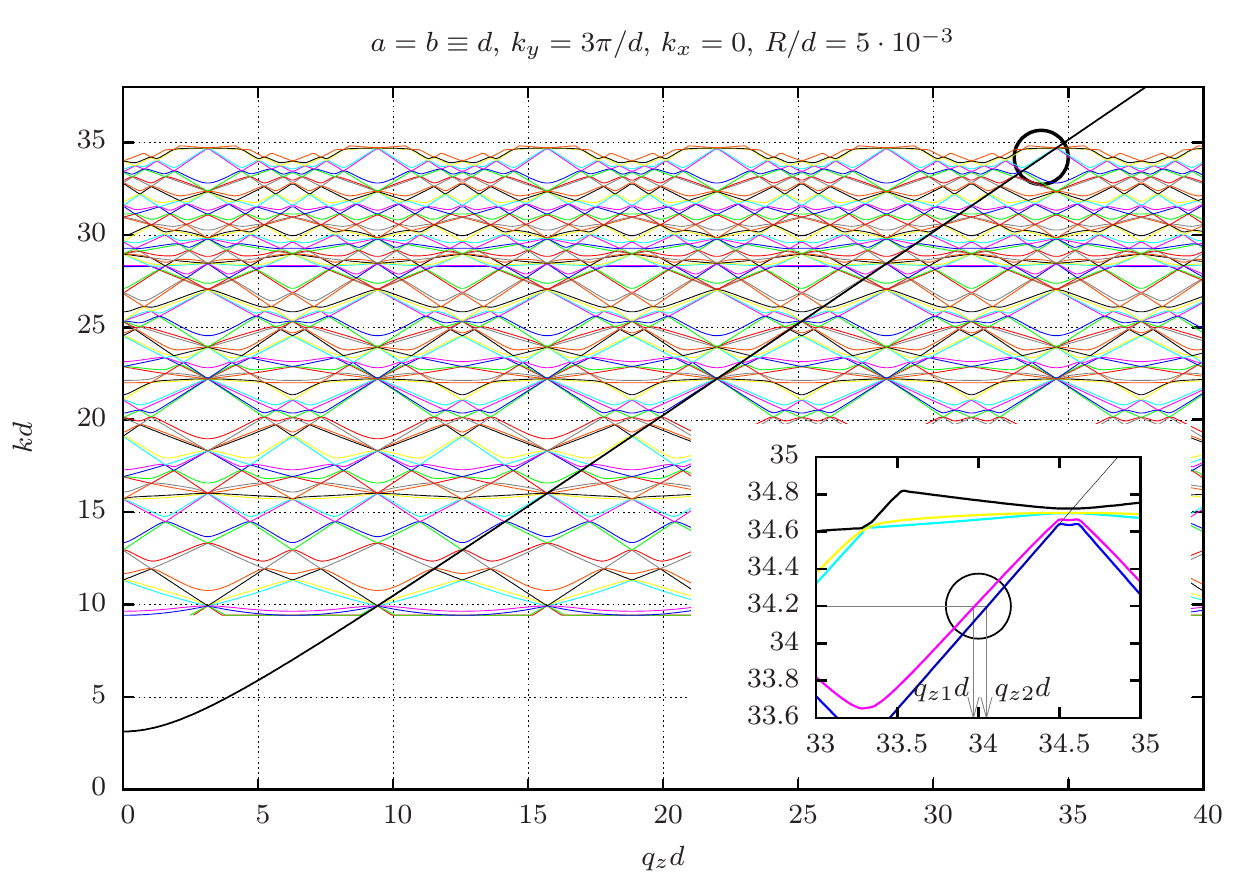}
    \includegraphics[width=0.47\linewidth]{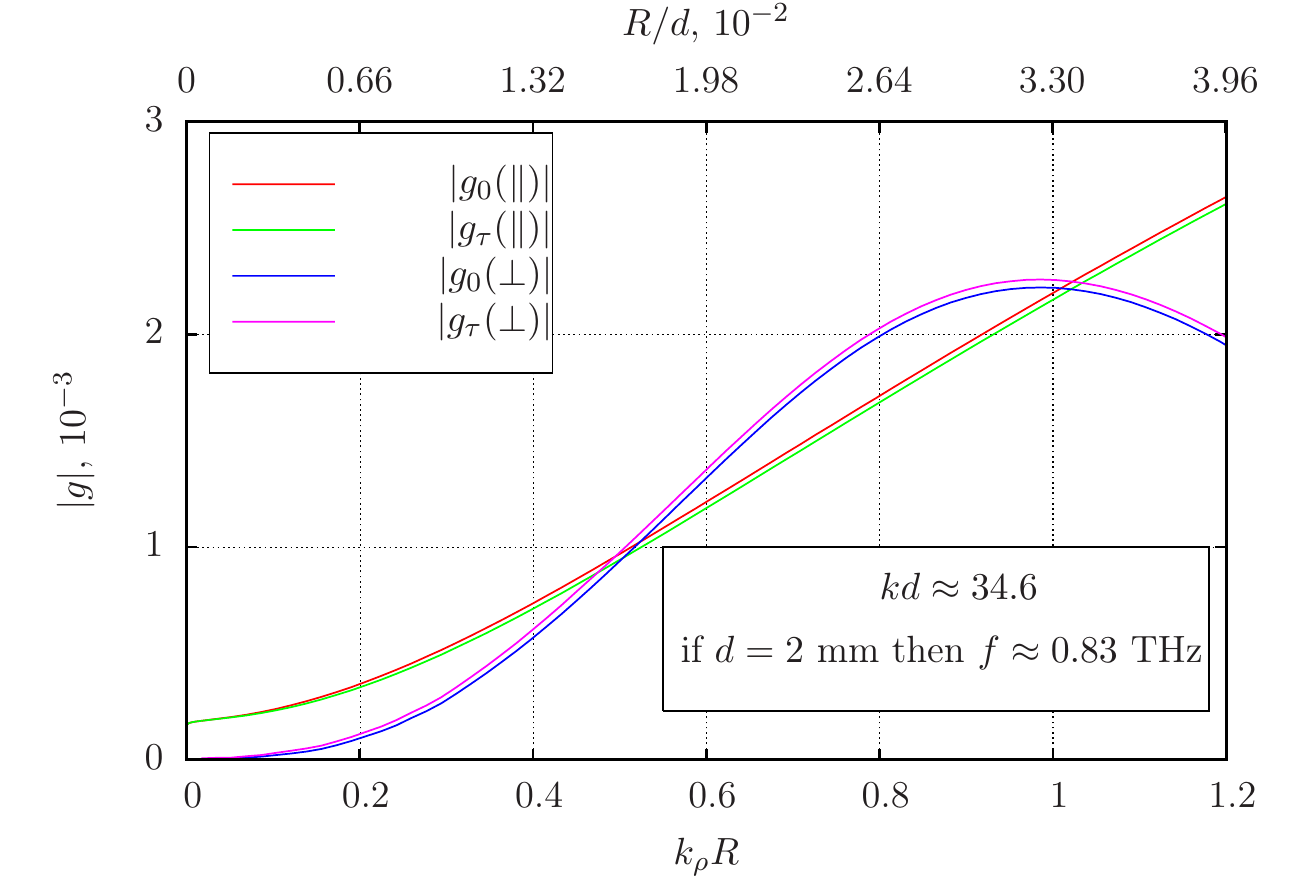}
    \caption{Left: dispersion curves for a crystal built from parallel metallic wires,
    placed in a vacuum ($\varepsilon_1=\mu_1=1$);
        Right: absolute values of effective polarizabilities $g_0$ and
        $g_\tau$ versus $k_\rho R$ for the selected values of the
        parameters}
    \label{fig:disp_curves}
\end{figure}
Figure~\ref{fig:disp_curves} gives an example of calculating $g_0$
and  $g_\tau$ using the dispersion equation \eqref{eq:DE_full}. We
consider a crystal with square lattice $a=b\equiv d$ and it
is assumed that $k_y=\frac{3\pi}{d}$ (i.e., at least the conditions
of symmetric two-wave diffraction in Laue geometry are fulfilled
in the crystal). The left graph shows the general view of the
dispersion curves for a TM-polarized wave at $R/d=5\cdot 10^{-3}$.
The black curve represents the solution of the dispersion equation
for vacuum ($k^2=k_z^2 + k_y^2$, $k_z\equiv q_z$).
 The magnified
image shows the range of high frequencies and indicates two roots
corresponding to the two close solutions of the dispersion
equations. The right graph shows the corresponding absolute values
of the effective polarizabilities calculated for the selected geometry
with varied parameter $k_\rho R$ in the range $0<k_\rho R<1.2$.
 It can be seen that for a TE-polarized wave, $g_0^\perp$ and
$g_\tau^\perp$ increase as $k_\rho R$ is increased, attaining the
maximum in the vicinity of $k_\rho R \sim 1$. For a TM-polarized
wave, the absolute values $|g_0^\parallel|$ and
$|g_\tau^\parallel|$ increase monotonically.
 Let us note here that
for a TM-polarized wave, the values of $g_0$ and $g_\tau$ are
negative, while for a TE-polarized wave they are positive (i.e.,
the refractive index for  a TE-wave is greater than unity).
In the considered case, the maximum values of $g_0^\perp$ and
$g_\tau^\perp$ even exceed the corresponding values
$|g_0^\parallel|$ and $|g_\tau^\parallel|$, though in the general
case, according to the calculations, the relation between these
values can change. However, it is important that in our
calculations, the behavior of effective polarizabilities remains almost
the same: for a TM-polarized wave $|g_0|$ and $|g_\tau|$ increase
monotonically as $k_\rho R$ is increased from  $0$ to $1$, while
for a TE-polarized wave $|g_0|$ and $|g_\tau|$ have a maximum at
large $k_\rho R$ ($k_\rho R \sim 1$; the exact position of the
maximum may shift within a narrow range.)

\section{Radiation in crystals built from wires at  $k_\rho R\lesssim 1$}

\subsection{Formulas for spectral-angular distribution}

The expressions \eqref{eq:Ex_field_crystal} for the fields  and
the system \eqref{eq:Phi_system} derived in the previous sections
enable the analysis of photon emission from a charged particle in
a crystal built from metallic wires in the case when $k_\rho R\sim
1$. Let a particle velocity $\mathbf{v}=(v_x, v_y, v_z)$; then the
particle trajectory $\mathbf{r}(t) = \mathbf{v}t$. Let us assume
for simplicity that $\varepsilon_1=\mu_1=1$. Substitution of
\eqref{eq:Ex_field_crystal} into the formula for spectral-angular
distribution of radiation \eqref{eq:Spec_ang_gen} followed by
certain transformation gives

\begin{equation}
\frac{d^2 N^s}{d\Omega d\omega} = \frac{e^2 Q^2}{\hbar c}\cdot
\frac{1}{\omega}\cdot |I^s|^2, \label{eq:spec_ang_crystal}
\end{equation}
where for  TM-polarization
\begin{multline}
I^\parallel = \frac{1}{k_\rho b}\sum\limits_{n}
\frac{1}{k_{zn}}\left\{ \left(\mathcal{F}_{0n}
+\mathcal{F}_{1n}\frac{k_{zn}}{k_\rho} +
\mathcal{F}'_{1n}\frac{k_{yn}}{k_\rho}\right)\frac{\mathbf{e}_x[\mathbf{k}_n,
    [\mathbf{k}_n, \mathbf{v}]]}{\omega - \mathbf{k}_n\mathbf{v}}\right. -\\-
\left.\left(\mathcal{F}_{0n}
-\mathcal{F}_{1n}\frac{k_{zn}}{k_\rho} +
\mathcal{F}'_{1n}\frac{k_{yn}}{k_\rho}\right)\frac{\mathbf{e}_x[\mathbf{k}_n^{(-)},
    [\mathbf{k}_n^{(-)}, \mathbf{v}]]}{\omega - \mathbf{k}_n^{(-)}\mathbf{v}}
\right\}, \label{eq:I_integral}
\end{multline}
for  TE-polarization
\begin{multline}
I^\perp = \frac{1}{k_\rho b}\sum\limits_{n}
\frac{1}{k_{zn}}\left\{ \left(\mathcal{F}_{0n}^\perp
+\mathcal{F}_{1n}^\perp\frac{k_{zn}}{k_\rho} +
\mathcal{F}^{'\perp}_{1n}\frac{k_{yn}}{k_\rho}\right)\frac{\mathbf{e}_x[\mathbf{k}_n,
    \mathbf{v}]}{\omega - \mathbf{k}_n\mathbf{v}}\right. -\\-
\left.\left(\mathcal{F}_{0n}^\perp
-\mathcal{F}_{1n}^\perp\frac{k_{zn}}{k_\rho} +
\mathcal{F}^{'\perp}_{1n}\frac{k_{yn}}{k_\rho}\right)\frac{\mathbf{e}_x[\mathbf{k}_n^{(-)},
    \mathbf{v}]}{\omega - \mathbf{k}_n^{(-)}\mathbf{v}}
\right\}, \label{eq:I_integral_perp}
\end{multline}
 $\mathcal{F}_{0n}$, $\mathcal{F}_{1n}$, and
$\mathcal{F}'_{1n}$ equal
\begin{equation}
\mathcal{F}_{0n} = \sum\limits_{m=1}^{M}F_{0m}e^{i \frac{z_m}{v_z}
(\omega - k_xv_x - k_{yn}v_y)},\;\; \mathcal{F}_{1n} =
\sum\limits_{m=1}^{M}F_{1m}e^{i \frac{z_m}{v_z} (\omega - k_xv_x -
k_{yn}v_y)},\;\; \mathcal{F}'_{1n} =
\sum\limits_{m=1}^{M}F'_{1m}e^{i \frac{z_m}{v_z} (\omega - k_xv_x
- k_{yn}v_y)}, \label{eq:SFcal}
\end{equation}
 where the
notations  $\mathbf{k}_n = (k_x, k_{yn}, k_{zn}) $ and
$\mathbf{k}_n^{(-)} = (k_x, k_{yn}, -k_{zn})$ are introduced.

Formulas \eqref{eq:spec_ang_crystal}-\eqref{eq:SFcal} together
with the set of equations \eqref{eq:Phi_system} describe the
emission from a charged particle passing through the crystal built
from wires.
Let us suppose that we have a photonic crystal consisting of a
small number of one-dimensional gratings (several tens or hundreds).
In this case, the set of equations
\eqref{eq:Phi_system} can be efficiently solved numerically. If
\eqref{eq:Phi_system} is successfully solved and the values of the
amplitudes $F_{0m}$, $F_{1m}$, and $F'_{1m}$ are found, then the
spectral-angular distribution of radiation can be
calculated using \eqref{eq:spec_ang_crystal}-\eqref{eq:SFcal}. But
since these expressions in the general form can hardly be
integrated analytically, we need to apply the numerical
integration when using them for calculating the total intensity of
radiation.

To simplify the analysis of the radiation process let us make use
of the parameters $g_0$, $g_{\pmb{\tau}}$ pre-calculated by the
dispersion equation. Let  the diffraction conditions in the
crystal be violated. If $|g_0|\ll 1$, then the wave vector in the
crystal $\mathbf{q}\approx \mathbf{k} + \frac{\omega g_0}{2 c
\gamma_0}\mathbf{N}$, where $\mathbf{N}$ is the normal to the
crystal surface and $\gamma_0=\mathbf{k}\mathbf{N}/k$. By solving
the boundary-value problem for plane-wave refraction by a crystal
plate of thickness $L$ placed at $0<z<L$ we can show that

\begin{equation}
\mathbf{E}_\mathbf{k}^{(-)s} \approx \mathbf{e}_s e^{i\mathbf{k}\mathbf{r}}\theta(z-L)
+ \mathbf{e}_s e^{i\mathbf{q}\mathbf{r}}e^{i\frac{\omega g_0 L}{2 c \gamma_0}}
\theta(L-z)\theta(z)
+ \mathbf{e}_s e^{i\mathbf{k}\mathbf{r}}e^{i\frac{\omega g_0 L}{2 c \gamma_0}}\theta(-z),
\label{eq:refract_solution}
\end{equation}
where $\theta(z)$ is the Heaviside function ($\theta(z)=1$ at
$z\geq 0$ and $\theta(z)=0$ at $z<0$). Substitution of
\eqref{eq:refract_solution} into  \eqref{eq:Spec_ang_gen} yields a
well-known expression for spectral-angular distribution of
Cherenkov and transition radiations:
\begin{equation}
\label{eq:cherenkov_trans}
\frac{d^2N^s}{d\omega d\Omega} = \frac{e^2Q^2\omega}{4\pi^2 \hbar c^3}(\mathbf{e}_{s}
\mathbf{v})^2\left| e^{i\frac{\omega L g_0^s}{2c\gamma_0}}
\left[\frac{1}{\omega-\mathbf{k}\mathbf{v}}-\frac{1}{\omega-\mathbf{q}\mathbf{v}}\right]
\left(e^{i(\omega-\mathbf{q}\mathbf{v})
\frac{L}{c\gamma_0}} -1\right)\right|^2.
\end{equation}

\begin{figure}[htp]
\centering
\begin{minipage}[h]{0.24\linewidth}
 \center{\includegraphics[height=5cm]{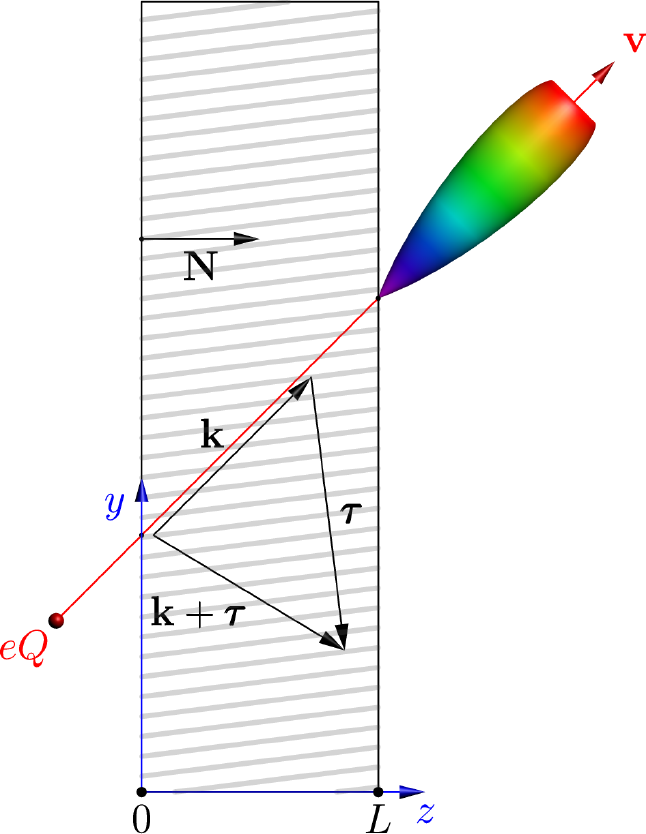} \\ a)}
\end{minipage}
\begin{minipage}[h]{0.24\linewidth}
 \center{\includegraphics[height=5cm]{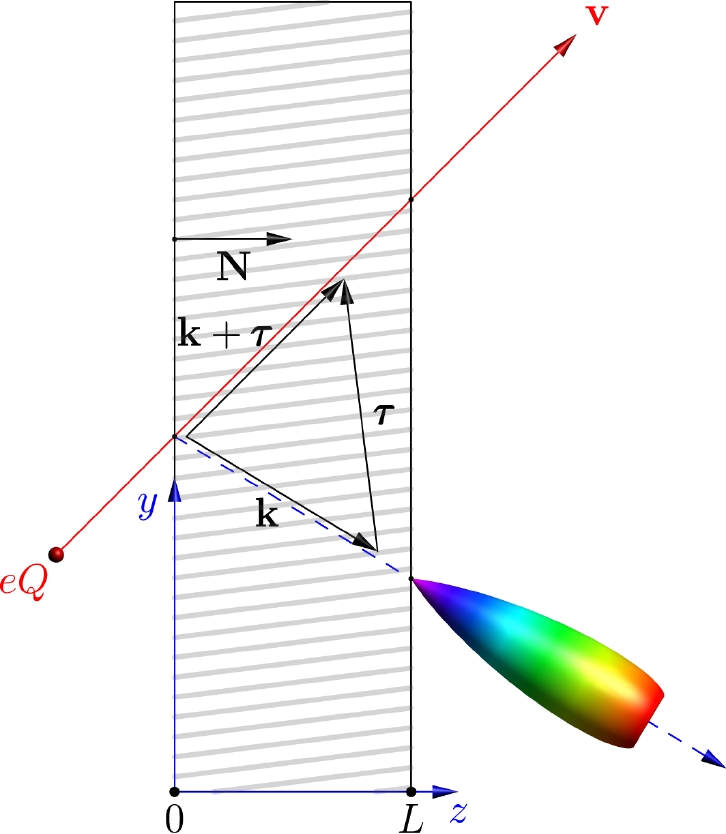} \\ b)}
\end{minipage}
\begin{minipage}[h]{0.24\linewidth}
 \center{\includegraphics[height=5cm]{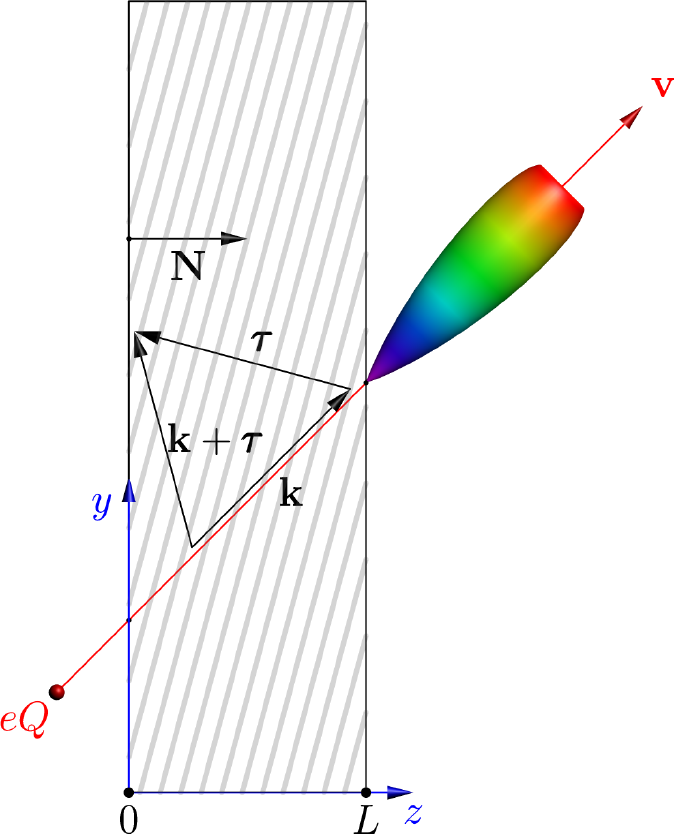} \\ c)}
\end{minipage}
\begin{minipage}[h]{0.24\linewidth}
 \center{\includegraphics[height=5cm]{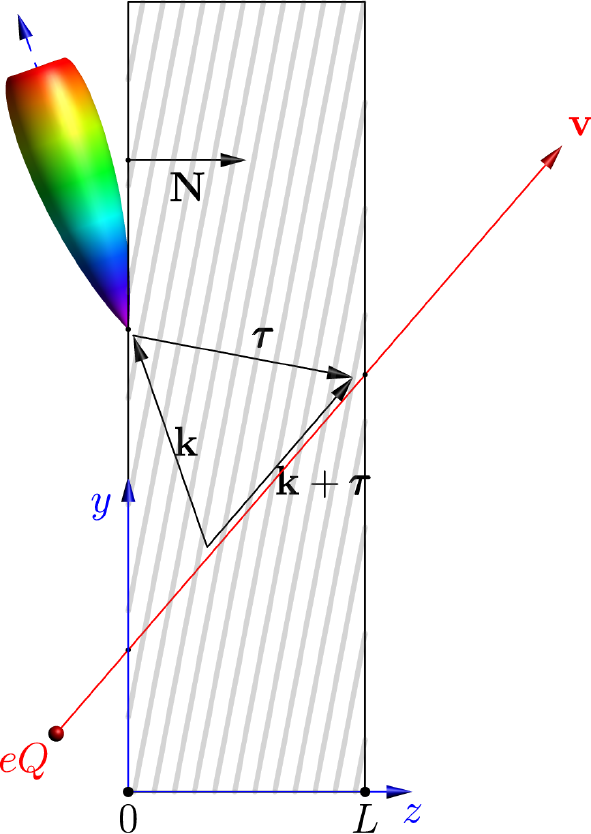} \\ d)}
\end{minipage}
\caption{Parametric (quasi-Cherenkov) radiation in photonic crystal for the case of two-wave diffraction
in Laue (a, b) and Bragg (c, d) geometries. The $x$-axis is perpendicular to the figure's plane.}
\label{fig:radiation_geometry}
\end{figure}

Now, let the two-wave diffraction conditions in the crystal be fulfilled.
Substitution of the formulas for $\mathbf{E}_\mathbf{k}^{(-)s}$,
which are valid in the case of diffraction, into
\eqref{eq:Spec_ang_gen} yields the expression for spectral-angular
distribution of parametric (quasi-Cherenkov) radiation (see Fig.
~\ref{fig:radiation_geometry}). Because the procedure is fully
described in \cite{Baryshevsky1997,Baryshevsky2015}, we shall not
be concerned discuss it here but shall be concerned with some
results only.

According to  \cite{Baryshevsky2015}, in the Laue case (when the
incident and diffracted waves leave the crystal through the same
surface) we have the following equation for parametric 
radiation at a small angle to particle velocity
(Fig.~\ref{fig:radiation_geometry}(a)):
\begin{eqnarray}
\label{eq:lauef} \frac{d^2N^s_0}{d\omega d\Omega} &=&
\frac{e^2Q^2\omega}{4\pi^2 \hbar c^3}(\mathbf{e}_{s}
\mathbf{v})^2\left| \sum_{\mu=1,2}\xi^0_{\mu
s}e^{i\frac{\omega}{c\gamma_0}\varepsilon_{\mu s}L}
\left[\frac{1}{\omega-\mathbf{k}\mathbf{v}}-\frac{1}{\omega-\mathbf{q}_{\mu
s}\mathbf{v}}\right] \left(e^{i(\omega-\mathbf{q}_{\mu
s}\mathbf{v}) \frac{L}{c\gamma_0}} -1\right)\right|^2,
\end{eqnarray}
and for radiation in the diffraction direction (Fig.~\ref{fig:radiation_geometry}(b))
\begin{eqnarray}
\label{eq:laued} \frac{d^2N^s_\tau}{d\omega d\Omega} &=&
\frac{e^2Q^2\omega}{4\pi^2 \hbar c^3}(\mathbf{e}^{\tau}_{s}
\mathbf{v})^2\left|\sum_{\mu=1,2}\beta_1\xi^\tau_{\mu
s}e^{i\frac{\omega}{c\gamma_0}\varepsilon_{\mu s}L}
\left[\frac{1}{\omega-\mathbf{k}_{\tau}\mathbf{v}}-\frac{1}{\omega-\mathbf{q}_{\tau\mu
s}\mathbf{v}}\right] \left(e^{i(\omega-\mathbf{q}_{\tau\mu
s}\mathbf{v}) \frac{L}{c\gamma_0}} -1\right)\right|^2,
\end{eqnarray}
where $\mathbf{q}_{\mu s}$ are the roots of the dispersion
equation, $\mathbf{k}_\tau = ((\mathbf{k}+\pmb{\tau})_\perp,
\sqrt{k^2-(\mathbf{k}+\pmb{\tau})_\perp^2})$, $\mathbf{q}_{\tau
\mu s} = \mathbf{q}_{\mu s}+\pmb{\tau}$, $\varepsilon_{\mu s} =
\gamma_0^2(q_{z\mu s}/k_z - 1)$,
$\gamma_0=\mathbf{k}\mathbf{N}/k$,
$\gamma_1=(\mathbf{k}+\pmb{\tau})\mathbf{N}/|\mathbf{k}+\pmb{\tau}|$,
$\beta_1=\gamma_0/\gamma_1$, $\mathbf{N}$ is the normal to the
entrance surface of the crystal directed to the crystal's
interior, the subscript ``$\perp$'' denotes the vector components
perpendicular to  $\mathbf{N}$, $\mathbf{e}_1||[\mathbf{k}\times
\pmb{\tau}]$, $\mathbf{e}_2||[\mathbf{k}\times \mathbf{e}_1]$, and
\begin{equation*}
\xi^0_{1(2)s} = \mp
\frac{2\varepsilon_{2(1)s}-g_0^s}{2(\varepsilon_{2s}-\varepsilon_{1s})},\;\;\;
\xi^\tau_{1(2)s} = \pm
\frac{g_\tau^s}{2(\varepsilon_{2s}-\varepsilon_{1s})}.
\end{equation*}
Similar expressions are obtained in \cite{Baryshevsky2015} for parametric radiation in
the Bragg diffraction case, too (when the incident and diffracted
waves leave the crystal through the opposite surfaces, Fig.~\ref{fig:radiation_geometry}(c--d)).

\subsection{Cherenkov and  transition radiations}
\label{subsec:Cherenkov} Now let us give a more detailed
consideration to different radiation cases that occur
as charged particles  moves uniformly through a crystal built from
metallic wires. Let us state, first, that because the crystal's
refractive index for a $TE$-polarized wave is greater than unity,
Cherenkov radiation is emitted in the crystal \cite{Gurnevich2009,
Gurnevich2010}. Moreover, transition radiation is also emitted
when a charged particle crosses the ``crystal-vacuum'' boundary. In
our analysis we shall use \eqref{eq:cherenkov_trans} for
spectral-angular distribution of transition and Cherenkov
radiations.

\begin{figure}[htp]
    \centering
    \includegraphics[width=0.7\linewidth]{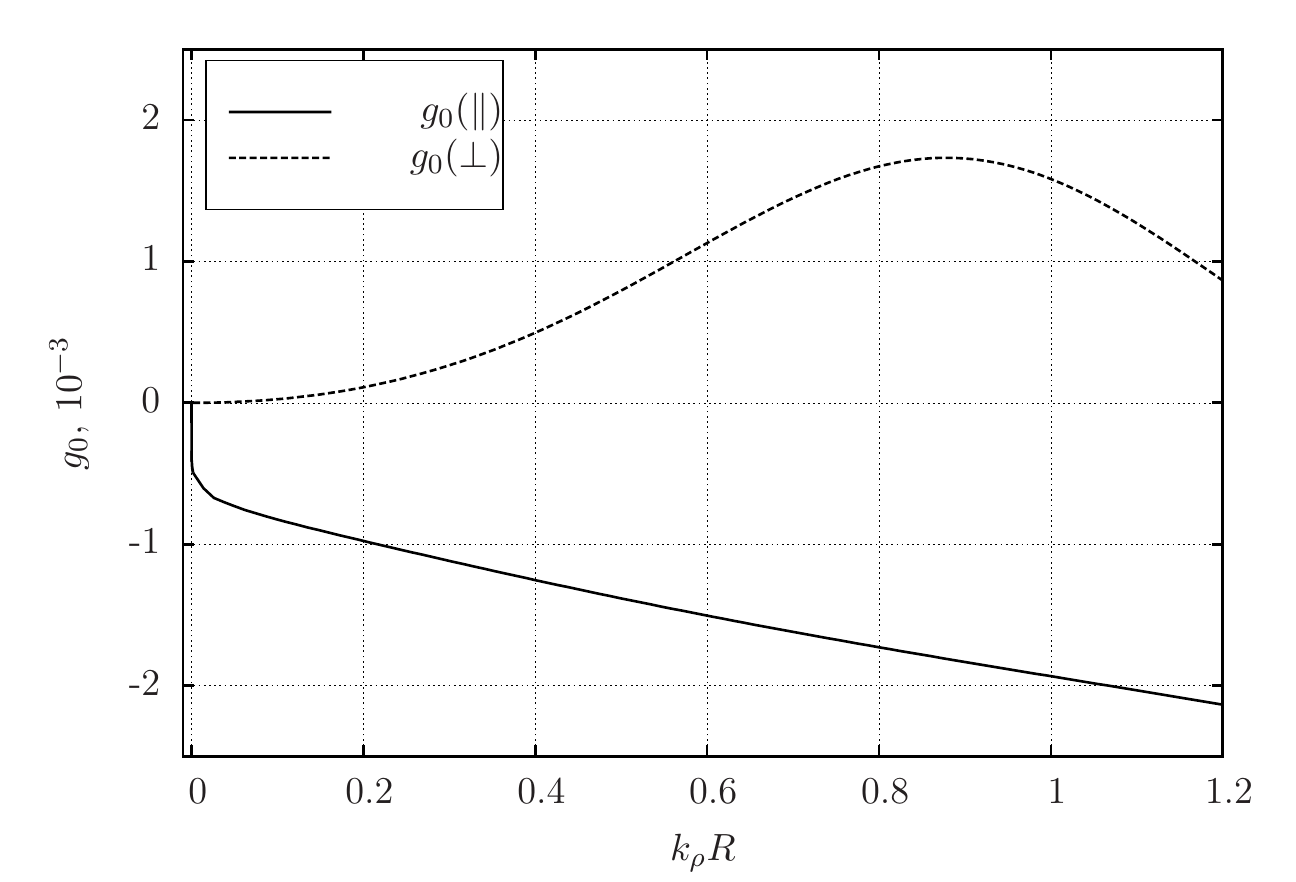}
    \caption{Effective polarizability $g_0$ as a function of the parameter  $k_\rho
    R$ for crystal made from metallic wires with a square lattice
    (normal incidence;  $kd\approx 35.5$, which at  $d=2$~mm
    corresponds to the frequency $0.85$~THz)}
        \label{fig:chi0_normal}
\end{figure}

For simplicity, we shall assume that the crystal has a square
lattice ($a=b\equiv d$). Figure~\ref{fig:chi0_normal} shows the
typical values of $g_0^\perp$ è $g_0^\parallel$  as a function of
the parameter $k_\rho R$  calculated by \eqref{eq:DU_normal}
 for a crystal made from metallic wires.

As is seen, for a $TE$-polarized wave, $g_0>0$, i.e., the crystal's
refractive index for this wave is $n_\perp\approx 1 +
\frac{1}{2}g_0^\perp>1$. If the particle velocity is less than
than the phase velocity  $v_\perp=c/n_\perp$ of a $TE$-wave in
crystal, then only transition radiation is emitted as the particle
passes through the crystal. But if the particle velocity is
greater than  the phase velocity of a $TE$-wave, then $\omega -
\mathbf{q}\mathbf{v}$ can vanish, and the term in
\eqref{eq:cherenkov_trans} whose denominator contains $\omega -
\mathbf{q}\mathbf{v}$ will increase as the crystal thickness is
increased. At significantly large $L$ ($L\gg l_0=\lambda\gamma^2$;
$l_0$ ($l_0$ is the coherent radiation length in the vacuum and
$\lambda$ is the wavelength), this term will make the major
contribution to the total radiation intensity.  This picture fully
corresponds to the ordinary Cherenkov radiation emitted as the
charged particle moves through optically transparent medium at a velocity
greater than the phase velocity of light for this media. As is
known, this radiation is emitted at an angle $\theta$ to the
direction of the particle velocity determined from the
condition $\cos \theta=\frac{1}{\beta n_\perp}$, where
$\beta=v/c$. In the ultra-relativistic case, when  $\gamma\gg 1$,
in view of smallness of $g_0^\perp$ the Cherenkov cone angle for
the considered crystal can be written as  $\theta_c\approx
\sqrt{g_0^\perp}$.

\begin{figure}[htp]
    \centering
    \includegraphics[width=0.99\linewidth]{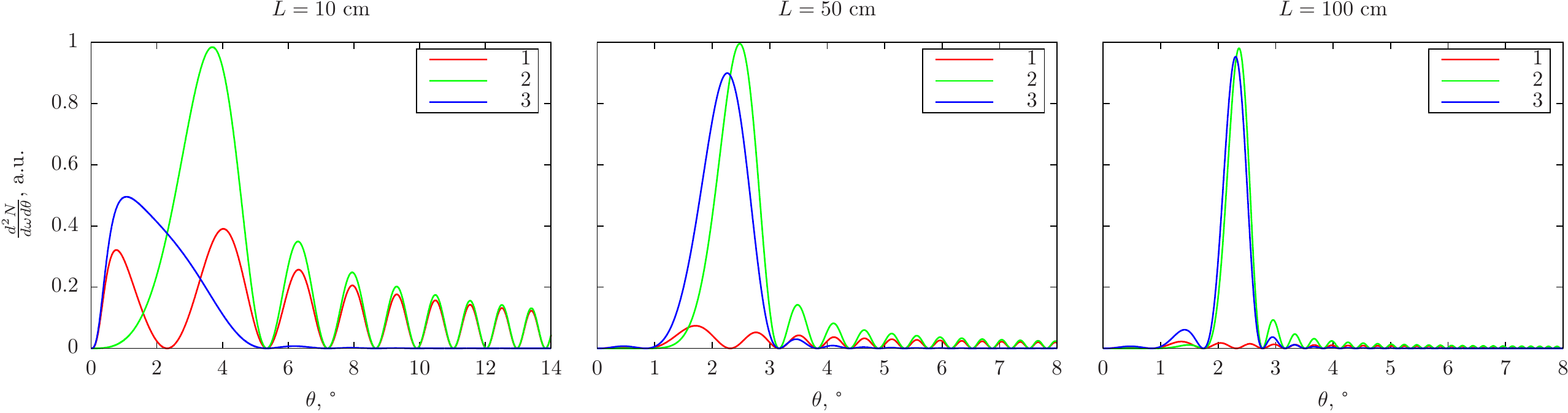}
    \caption{Spectral-angular distributions of Cherenkov and transition radiations
    in a crystal made from metallic wires for a $TE$-polarized wave:
    1 -- only the first term between the square brackets in \eqref{eq:cherenkov_trans} is taken into account,
     2 -- only the second term between the square brackets in \eqref{eq:cherenkov_trans} is taken into
     account,
    3 -- all terms are taken into account. The given dependence is on the polar angle $\theta$
    alone     (the polarization factor  $\sin^2 \varphi$ is omitted).
    Particle Lorentz factor  $\gamma=100$,
    crystal period $d=2$~mm, radiation frequency $f=0.85$~THz,
    $l_0=\lambda \gamma^2\approx 3.5$~m,
    $g_0^\perp\approx1.7\cdot 10^{-3}$.}
    \label{fig:spec_ang_cher}
\end{figure}

It should be noted, however, that at typical values $\gamma\sim
10^2\div10^3$ and $\lambda\sim 0.3$~mm, the coherent length
 $l_0$ takes on rather large values in the range from several
  meters to several hundreds of meters, so in the general case, to find spectral-angular distribution,
  we need   to consider all terms appearing in \eqref{eq:cherenkov_trans}.
By way of example, Figure~\ref{fig:spec_ang_cher} compares the
spectral-angular distributions of Cherenkov and transition
radiations calculated considering all terms between the square
brackets in \eqref{eq:cherenkov_trans} with those calculated
considering only the first or only the second term. In our
calculations the particle Lorentz factor was $\gamma=100$, only
radiation of $TE$-polarized wave was considered, and the
assumed value of $g_0^\perp\approx1.7\cdot 10^{-3}$ corresponded
to  $k_\rho R \approx 0.9$ (see Fig.~\ref{fig:chi0_normal}). As
is seen, with the selected values of these parameters, the results
obtained using the complete formula \eqref{eq:cherenkov_trans} and
those obtained taking into account only the term proportional to
$(\omega - \mathbf{q}\mathbf{v})^{-1}$ are almost the same when
the crystal thickness $L\gtrsim 1$~m.

For a $TM$-polarized wave, $g_0<0$ (see
Fig.~\ref{fig:chi0_normal}), and the denominator of the second term
in \eqref{eq:cherenkov_trans} cannot vanish. In this case only
transition radiation is possible in the crystal; its
spectral-angular distribution is analyzed with due account of all
terms in \eqref{eq:cherenkov_trans}.

\begin{figure}[htp]
    \centering
    \includegraphics[width=0.7\linewidth]{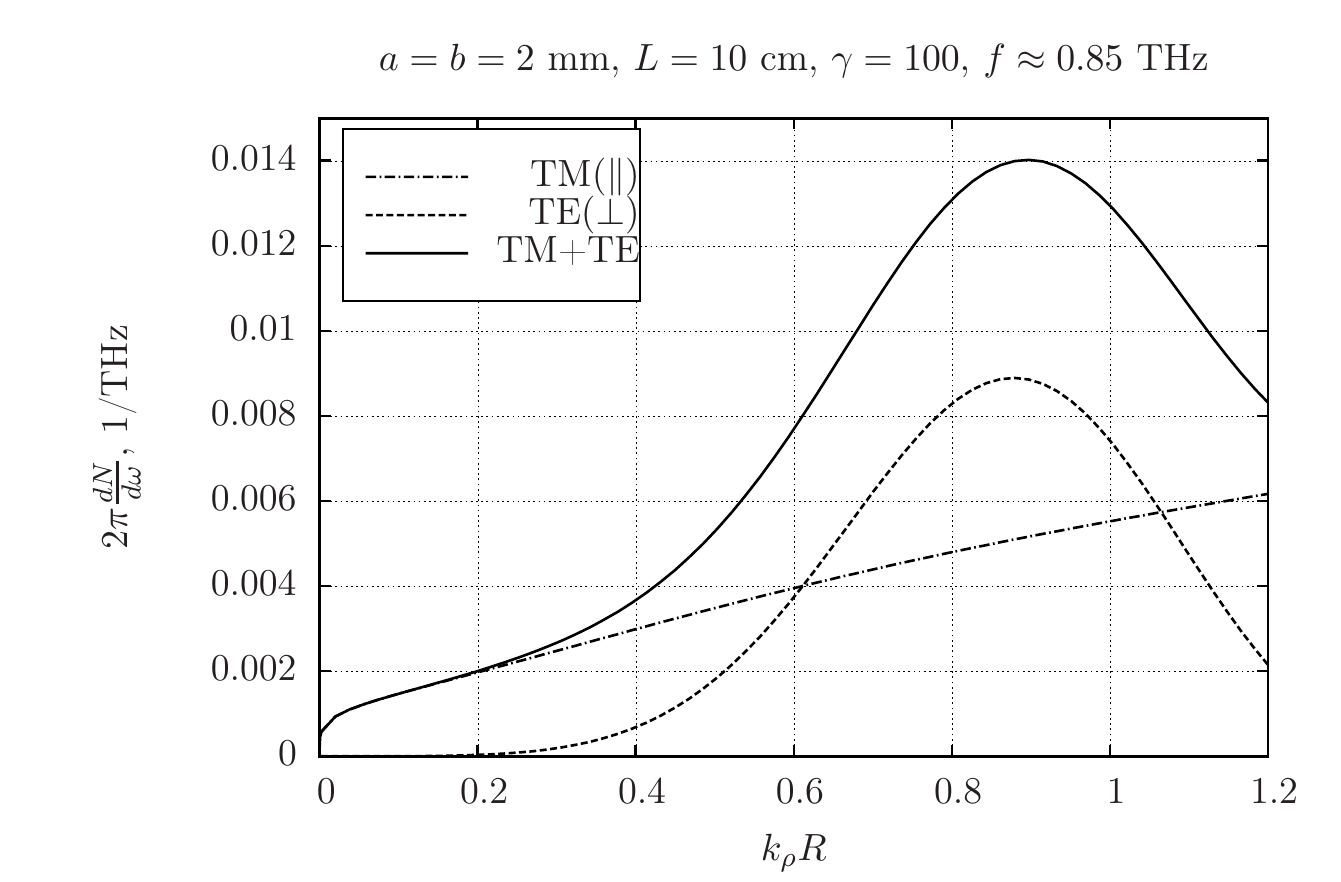}
    \caption{Intensities of transition and Cherenkov radiations in the crystal formed by metallic wires as
    a function of  $k_\rho R$.
   Radiation intensity from $TM$ and $TE$-polarized waves and total radiation intensity (for both polarizations) are
   shown. The results are obtained by formula
     \eqref{eq:cherenkov_trans} with due account of all terms.}
    \label{fig:dNdw}
\end{figure}

The total radiation intensity (without separation of contributions
coming from Cherenkov and transition radiations) can be found by
numerical integration of \eqref{eq:cherenkov_trans} with respect
to angular coordinates and frequency. By way of example we
calculated the intensities of Cherenkov and transition radiations
in a crystal built from metallic wires using the known values of
$g_0$ (Fig.~\ref{fig:chi0_normal}). To avoid ambiguity, the
crystal thickness was set equal to 10~cm, the particle (electron)
velocity was perpendicular to the crystal surface, $\gamma=100$.
The results are given in Fig.~\ref{fig:dNdw}.
As is seen in the plot, the intensity of transition and Cherenkov
radiations in a crystal made from metallic wires has a maximum in
the vicinity of $k_\rho R\sim 1$ that corresponds to the maximum
value of $g_0^\perp$. At $k_\rho R \approx 0.9$, in particular,
the total radiation intensity for two polarizations is as high as
$1.4\cdot10^{-2}$~photons/THz at the frequency $f=0.85$~THz. Then,
for example, in a narrow frequency range from $f_1=0.99f$ to
$f_2=1.01f$ ($\Delta f = 0.02f=17$~GHz) the number of photons
emitted per one electron is on average $N_{tot}\sim
2.4\cdot10^{-4}$~photons. Of course, in a wider frequency range,
the total number of photons emitted per electron will be greater,
achieving , e.g., $N_{tot}\sim 2.4\cdot10^{-3}$ at $\Delta f =
0.2f=170$~GHz. Note here that the intensity of radiation, being
 proportional to $Q^2$ according to
\eqref{eq:cherenkov_trans}, is much higher for particles with
greater charge ($Q>1$). Thus, for a relativistic nucleus of charge
$Q\sim 50$ passing through the same crystal of thickness 10~cm,
the intensity of radiation at $k_\rho R \sim 1$ increases
by more than three orders of magnitude up to more than 1 photon
per nucleus \cite{Baryshevsky2015}.

\FloatBarrier

\subsection{Parametric (quasi-Cherenkov) radiation}
\label{subsec:quasi-cherenkov}

We shall proceed to the discussion of parametric radiation in
a crystal made from metallic wires. For definiteness, we shall
consider radiation in the case of two-wave symmetric Laue
diffraction (see Fig.~\ref{fig:radiation_geometry}a-b). Here the
reciprocal lattice vector is $\pmb{\tau}=\frac{2\pi
m}{d}\mathbf{e}_y$, where $m$ is the integer, and the Bragg angle
$\theta_B$ is related to $k_B$ as
\begin{equation}
 \sin \theta_B = \frac{\tau}{2k_B}=\frac{\pi m}{kd}.
\end{equation}
By way of example we assume that $m=3$ and make use of the above
values of the effective polarizabilities $g_{\pmb{\tau}}$
(Fig.~\ref{fig:disp_curves} right). We also assume that the
particle has a charge
 $Q=1$ (electron, proton, etc.).
The intensity of radiation is calculated using \eqref{eq:laued}.

\begin{figure}[htp]
    \centering
    \includegraphics[width=0.7\linewidth]{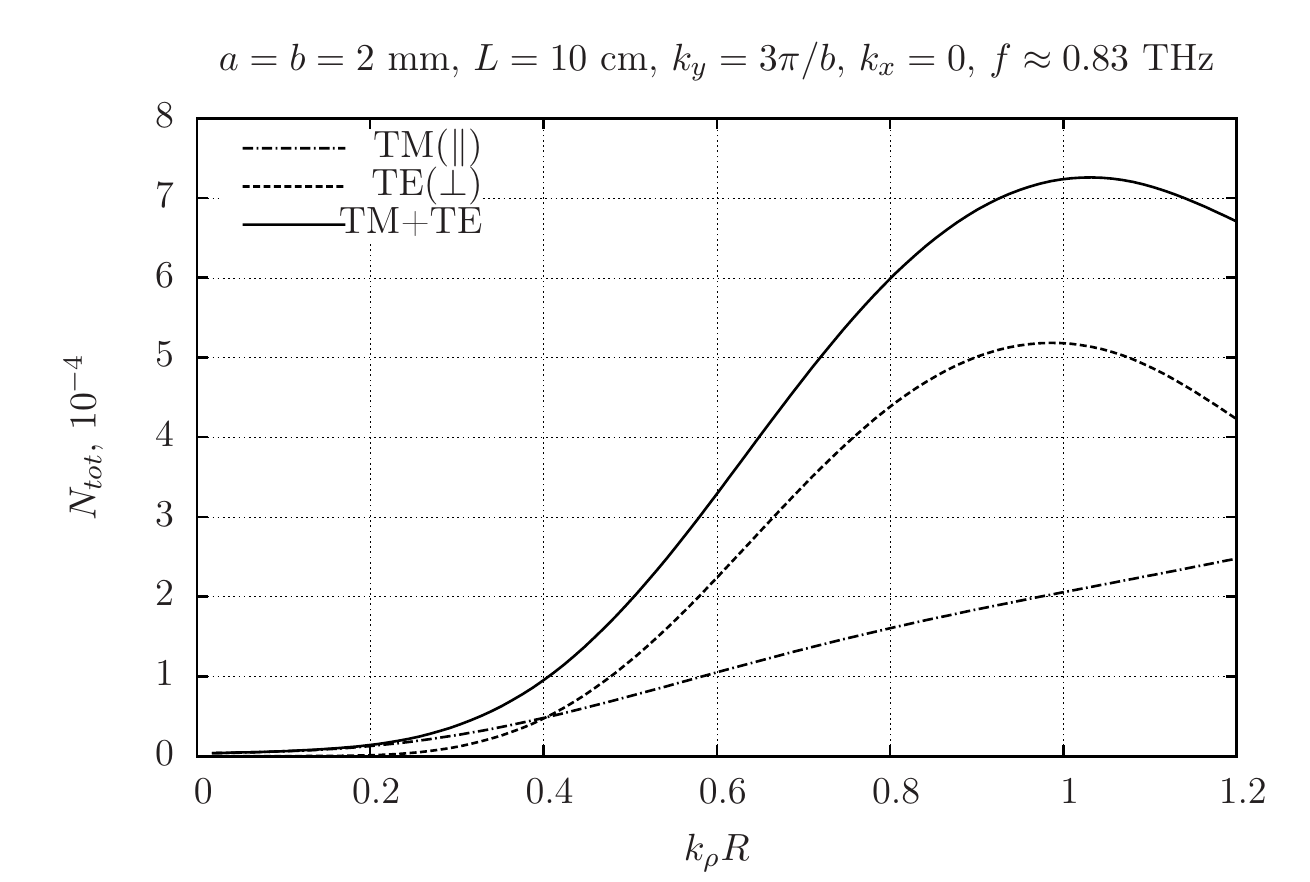}
    \caption{Intensity of parametric radiation (number of photons
        emitted by one electron passing through the crystal) as a
        function of radius of the wires composing the crystal.}
    \label{fig:full_intensity}
\end{figure}

The results of numerical integration of these expressions with
respect to frequency and angular coordinates are given in
 Fig.~\ref{fig:full_intensity}. As is seen, the radiation intensity for $TM$-polarization
 increases monotonously with $k_\rho R$, attaining at $k_\rho R =
 1.2$ the value $N_{\parallel}\approx
2.5\cdot 10^{-4}$ photons/electron (the radiation frequency for
the selected crystal period of 2~mm is $0.83$~THz). At large
$k_\rho R$, the intensity of radiation for a $TE$-polarized wave
also increases appreciably, in the considered case exceeding that
for $TM$-polarization already at $k_\rho R \gtrsim 0.4$ and
achieving the maximum value $N_{\perp}\approx 5\cdot 10^{-4}$
photons/electron at $k_\rho R\approx 1$.
The total (for two polarizations) intensity of radiation also
attains its maximum $N_{tot}\approx 7\cdot
10^{-4}$~photons/electron at $k_\rho R\approx 1$ (see
Fig.~\ref{fig:full_intensity}), which for the selected parameters
of the crystal corresponds to the wire radius $R\approx 65~$~$\mu$m.

\begin{figure}[htp]
    \centering
    \includegraphics[width=0.99\linewidth]{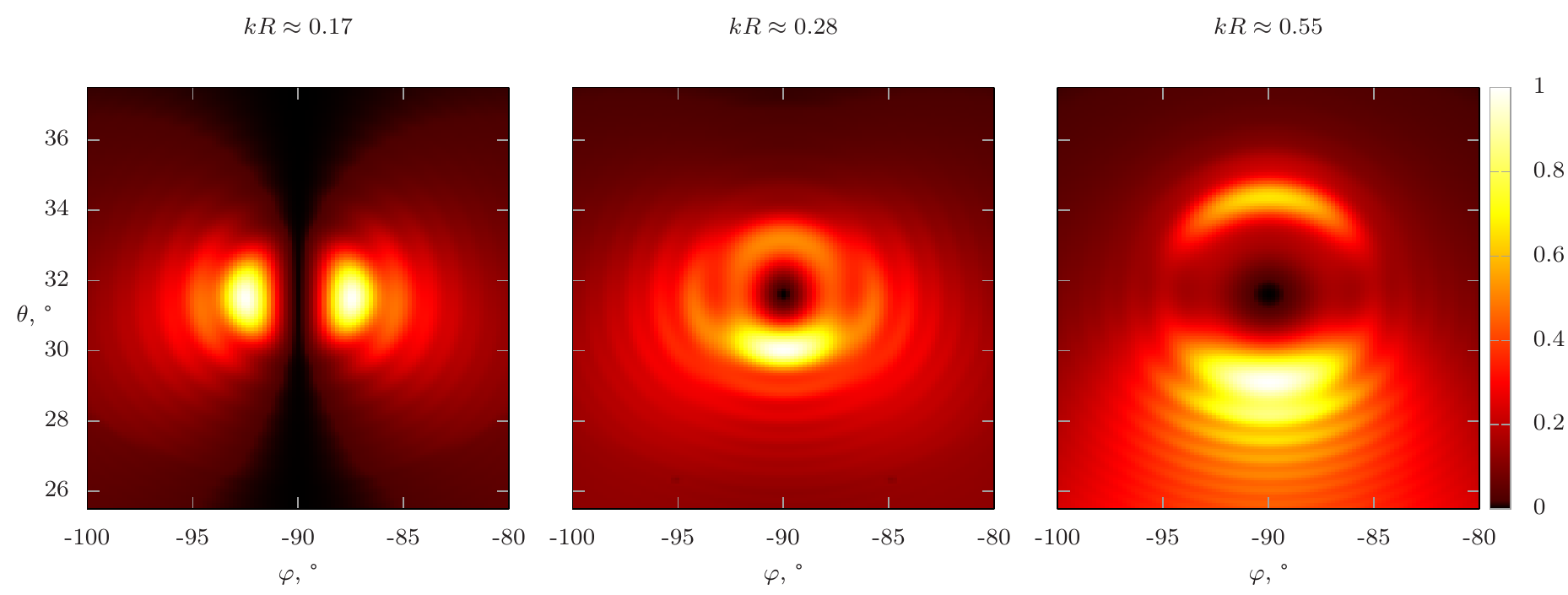}
    \caption{Angular distribution (total for two polarizations) of
        parametric radiation in the diffraction direction at different   $k_\rho R$;
        the crystal thickness is $L=50$~cm, Lorentz factor $\gamma=100$ (electron energy
        $E=50.5$~MeV).}
    \label{fig:ang_full}
\end{figure}

Let us pay attention to the fact that at large $k_\rho R$, the
contributions to the total intensity coming from $TE$- and
$TM$-polarized waves can be comparable (in our case at $k_\rho
R\sim 1$ the contribution from the $TE$ wave is dominating),
whereas for thin wires ($k_\rho R\ll 1$), the main contribution to
the radiation intensity comes from the $TM$-polarized wave. This
is also seen in Fig.~\ref{fig:ang_full} showing the angular
distribution of parametric radiation. As the parameter
$k_\rho R$ increases, the angular distribution  changes
appreciably from the form shown in Fig.~\ref{fig:ang_full} (left) at
$k_\rho R\ll 1$ to a more symmetric form (Fig.~\ref{fig:ang_full},
center) as the intensities of $TE$- and $TM$-polarized waves
become comparable, and finally to a form  shown in
Fig.~\ref{fig:ang_full} (left) as the main generation begins at
the $TE$-wave. Let us note that at large Bragg angles, the
$TE$-polarized wave can make a significant contribution to the
radiation intensity even when $k_\rho R\ll 1$, because
$A_0^{\perp}$ is nonzero (this is what makes the considered case
different from  parametric X-ray radiation where at
$2\theta_b=90^\circ$  the $\pi$-polarized is absent at all
\cite{PXR})


\begin{figure}[htp]
    \centering
    \includegraphics[width=0.5\linewidth]{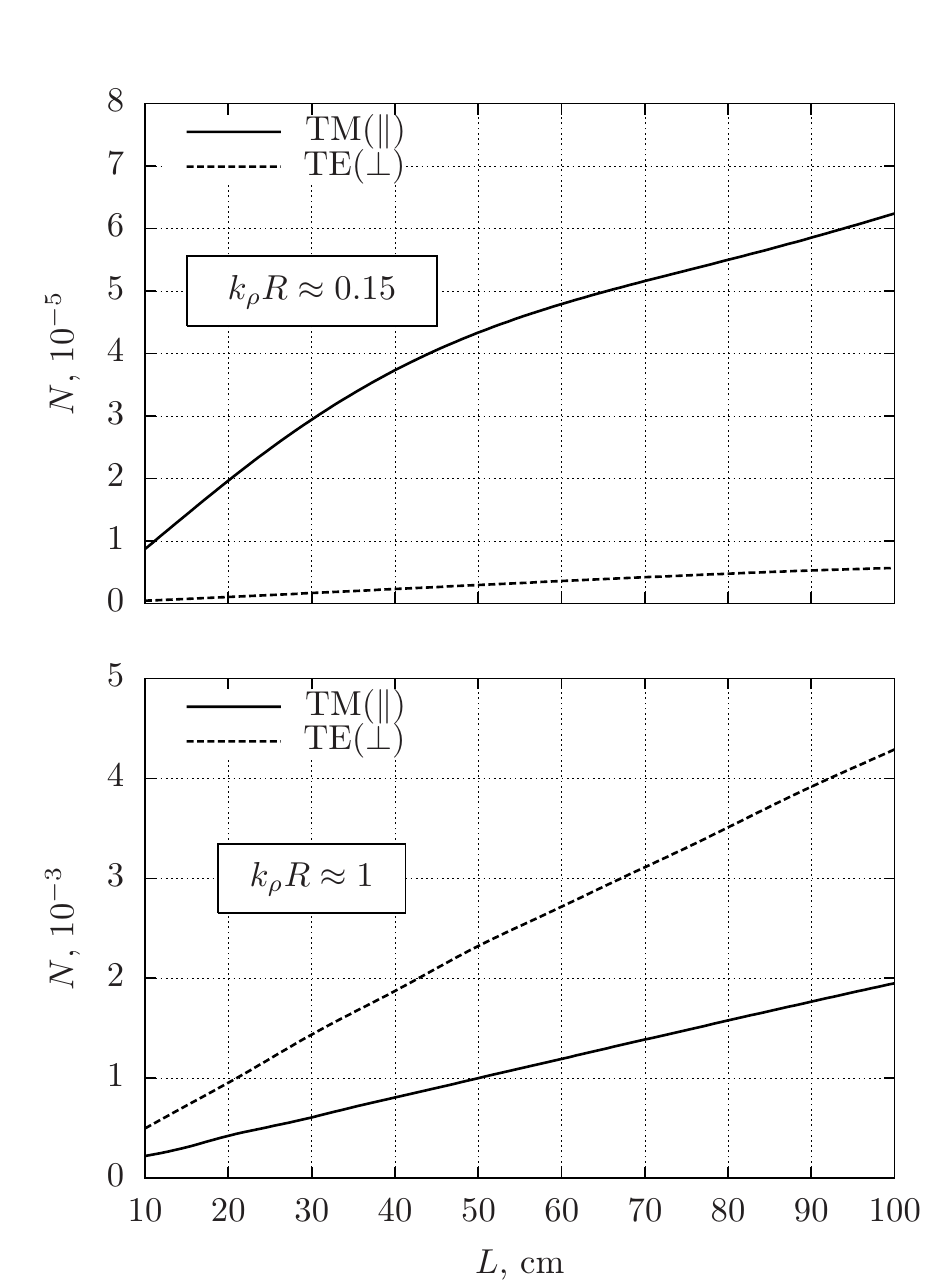}
    \caption{Intensity of parametric radiation as a function of crystal thickness.}
    \label{fig:full_N_L}
\end{figure}

As the photon absorption length $L_{abs}$ in the considered
crystal is large (e.g., in the case of copper wires, the
calculation by the dispersion equation gives $\Im g_0 \sim 8.5
\cdot 10^{-7}$, which corresponds to $L_{abs}\sim 10^2$~m at the
frequency $0.83$~THz), the total radiation intensity in the
crystal can be increased by increasing the crystal thickness. As
an example,  Fig.~\ref{fig:full_N_L} plots the intensity of
parametric radiation against the crystal thickness at varied
$k_\rho R$ for the selected geometry. It is seen that as $L$
increases the radiation intensity also increases, being as high as
$6\cdot 10^{-3}$ photons/electron  and greater at $k_\rho
R\sim 1$ and $L=1$~m.
In this case, at the selected parameters of the crystal the
intensity of radiation from the $TE$-polarized wave is almost
twice as large as that from the $TM$-polarized wave. Let us
emphasize that at small $k_\rho R$, almost the entire radiation is
generated at the $TM$-wave, which is obvious from
Fig.~\ref{fig:full_N_L}.

Let us take notice of the fact that in our discussion we did not
consider the influence of multiple scattering of particles in
crystals  on radiation process. This influence can be eliminated
if a particle moves through a hole made in the crystal or
parallel to the crystal surface. In this case, the radiation
processes are similar to those occurring in solid crystals,
provided that the distance $d$ between the particle and the
crystal surface satisfies the condition
$d<\frac{\lambda}{4\pi}\beta\gamma$
\cite{baryshevsky2012high,Bolotovsky}. In our example, we obtain
for $d$ a reasonable value $d\approx 3$~mm at $f=0.83$~THz and $\gamma=100$.

\FloatBarrier

\subsection{Radiation from electron bunches}

Beams of high-energy particles generated by accelerators usually
consist of short particle bunches following one another at certain
intervals. Modern acceleration facilities are capable of
generating relativistic electron beams with typical bunch duration
 $t_b$ less than  $10^{-13}$~s (and as low as
$10^{-15}$~s) with the number of electrons in the bunch   $N_e\sim
10^9$ and greater \cite{bunch_m1,bunch_m2,bunch3}. Such compact
bunches (whose size is much less than the wavelength) can emit
coherently (as a single particles of charge $Q=N_e$), and hence we
can expect that proceeding from single electrons to electron
bunches will significantly --- by a factor of  $N_e^2$ --- increase the
intensity of parametric radiation  \cite{Baryshevsky2015}.

According to \cite{Baryshevsky2011,Baryshevsky2015s}, the
spectral-angular distribution of photons emitted by a bunch of
particles moving in a crystal, $\frac{d^2N}{d\omega d\Omega}$,  is
related to that
 emitted by a single particle, $\frac{d^2N_1}{d\omega
d\Omega}$ as
 \begin{equation}
\frac{d^2N}{d\omega d\Omega}  = N_e \frac{d^2N_1}{d\omega d\Omega}
+ \frac{d^2N_1}{d\omega d\Omega}
\left| \int e^{-i \mathbf{K}\mathbf{r}}\rho(\mathbf{r})d^3 \mathbf{r} \right|^2,
\label{eq:sp_an_bunch_general}
\end{equation}
where $\rho(\mathbf{r})$is the bunch density and    $N_e = \int
\rho(\mathbf{r})d^3\mathbf{r}$ (integration is performed in the
entire bunch volume); the velocity spread in the bunch is
neglected. If the direction of the $z$-axis of the rectangular
coordinate system coincides with the direction of bunch velocity
 $\mathbf{v}$, then vector  $\mathbf{K}$
in \eqref{eq:sp_an_bunch_general}  for the forward parametric
radiation  $\mathbf{K}=(\mathbf{k}_\perp, \omega/v)$, whereas for
radiation in the direction of diffraction
$\mathbf{K}=((\mathbf{k}+\pmb{\tau})_\perp, \omega/v)$, where the
subscript  ``$\perp$'' is for vector components  perpendicular to
the  $z$-axis.

In many cases the density of particles in a bunch can be described
with good accuracy by normal distribution
\begin{equation}
\rho(\mathbf{r}) = \frac{N_e}{(\sqrt{2\pi})^3\sigma_\perp^2 \sigma_z}
\exp\left(-\frac{\mathbf{r}_\perp^2}{2\sigma_\perp^2}\right)
\exp\left(-\frac{z^2}{2\sigma_z^2}\right),
\label{eq:bunch_density}
\end{equation}
where the root-mean-square deviations  $\sigma_\perp$ and
$\sigma_z$ determine the bunch dimensions in transverse and
longitudinal directions (relative to the velocity direction),
respectively. Such bunches are used, e.g., for producing radiation
in free electron lasers \cite{ganter2010swissfel}. Substitution of
the expression for  $\rho(\mathbf{r})$ into
\eqref{eq:sp_an_bunch_general} gives
\begin{equation}
\frac{d^2N}{d\omega d\Omega}  = N_e \frac{d^2N_1}{d\omega d\Omega}
+ N_e^2 \frac{d^2N_1}{d\omega d\Omega}
\exp\left(-K_\perp^2\sigma_\perp^2 - \frac{\omega^2}{v^2}\sigma_z^2\right).
\label{eq:sp_an_bunch_gauss}
\end{equation}

By way of example, let us estimate the radiated power of
parametric radiation from a relativistic electron bunch passing
through a considered crystal made from metallic wires. We shall
use the typical bunch parameters  available with modern
acceleration facilities. For example, according to
\cite{ganter2010swissfel}  SwissFEL, X-ray free-electron laser
currently being built at the Paul Scherrer Institute can generate
electron beams composed of bunches with $N_e\approx 1.25\cdot
10^9$ (the bunch charge $Q\approx 200$~pC), $\sigma_z \approx
9$~$\mu$m (corresponds to the bunch duration of 30~fs),
$\sigma_\perp \approx 80$~$\mu$m. The plots given in Fig.~\ref{fig:ang_full}
show that the width of angular distributions
of parametric radiation in the selected geometry is within
$10^\circ$. From this we can readily estimate the maximum value
 $K_{\perp
max}\approx \frac{\omega}{c} \sin 10^\circ$ and the minimum value
of the exponential factor in the second term in
\eqref{eq:sp_an_bunch_gauss}: $\exp(-K_\perp^2\sigma_\perp^2 -
\frac{\omega^2}{v^2}\sigma_z^2)\approx 0.92$ (we considered here
that the radiation frequency in the example given in section~\ref{subsec:quasi-cherenkov}
is $f=\omega/2\pi\approx0.83$~THz).
Thus, the main contribution to the total intensity comes from the
second term in \eqref{eq:sp_an_bunch_gauss}, i.e., the electrons
in the bunch emit coherently. Then the instantaneous (peak) power
of parametric radiation is $P\approx 0.92\frac{N_{tot}N_e^2 \hbar
\omega}{t_b}$, where $t_b=\sigma_z/c\approx 30$~fs is the bunch
duration. Substitution of the found maximum value for the crystal
of thickness  10~cm $N_{tot}\approx 7\cdot 10^{-4}$, gives
$P\approx18$~MW. Using Fig.~\ref{fig:full_N_L} we can readily find
that as the crystal thickness increases to 1~m, the radiation
power increases by almost a factor of 10, achieving the value as
high as $P\approx 160$~MW.

For comparison, let us estimate the power of transition and
Cherenkov radiations. For parametric radiation, the width
of the spectral peak  $\Delta\omega/\omega \sim \sqrt{\gamma^{-2}
+ |g_0|}\approx 0.04$ \cite{BarJPhysFrance1983}. Using the results
of section~\ref{subsec:Cherenkov}, we obtain that the electron
passing through the crystal of thickness 10 cm emits in the same
frequency range  on average  $N_{tot}\approx 1.4\cdot 10^{-2}\cdot
0.04 = 5.6\cdot10^{-4}$ photons of Cherenkov and transition
radiations. Then the appropriate instantaneous power of radiation
produced by the considered electron bunch is about  14~MW (for
1m-thick crystal, the estimated value is  $P\approx 130$~MW).

So we can conclude that the intensity of transition and Cherenkov,
as well as parametric (quasi-Cherenkov) radiations in the THz
range for a crystal built from metallic wires is sufficient not
only for experimental observations but also for possible
applications \cite{Baryshevsky2015}, e.g. for development of
high-power THz sources.

\begin{figure}[h]
    \begin{minipage}[h]{0.39\linewidth}
        \center{\includegraphics[width=0.99\linewidth]{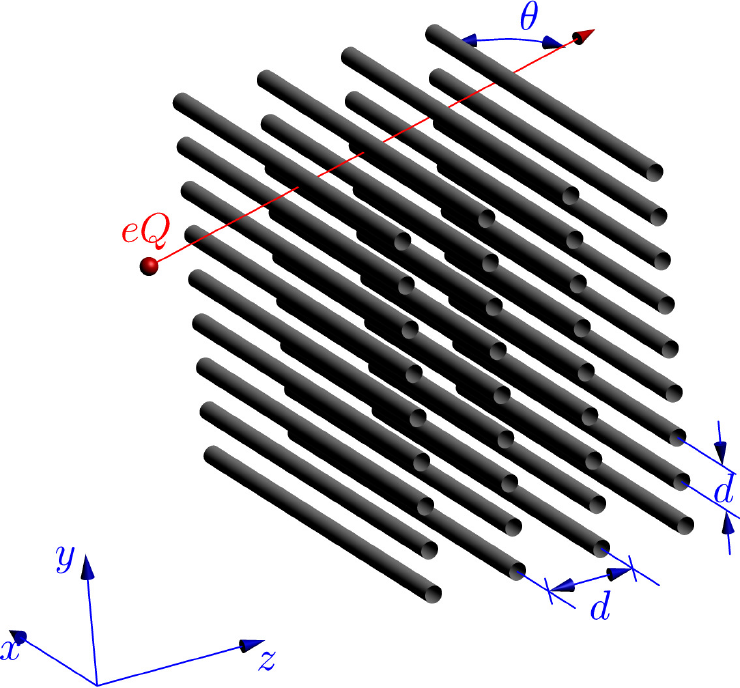} \\a)}
    \end{minipage}
    \begin{minipage}[h]{0.60\linewidth}
        \center{\includegraphics[width=0.99\linewidth]{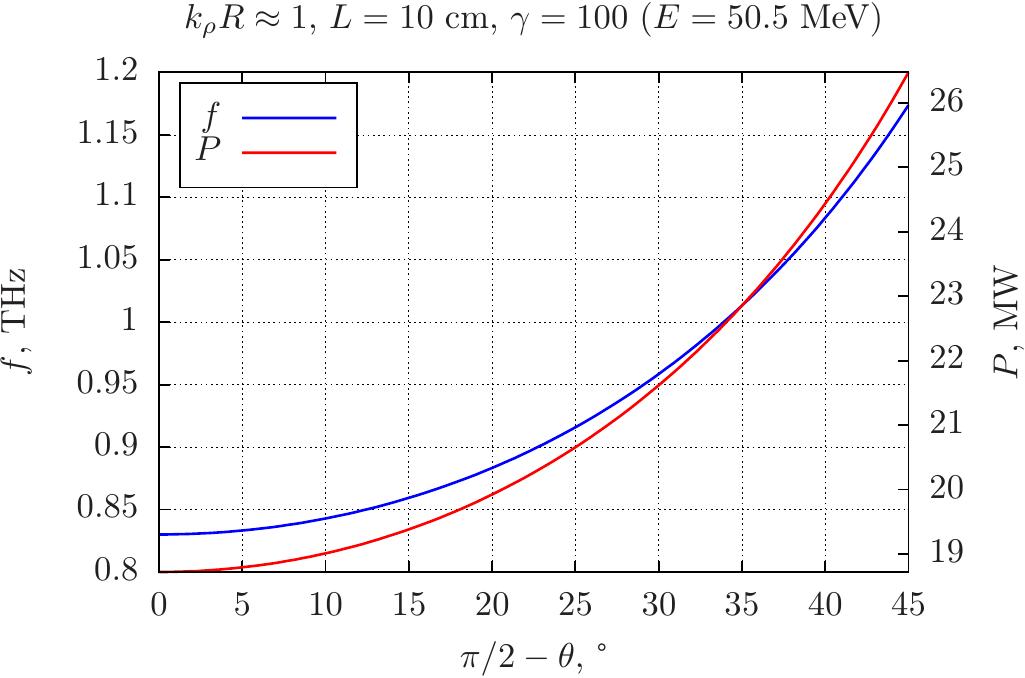} \\b)}
    \end{minipage}
    \caption{The geometry of the problem (a) and dependencies of the parametric 
    radiation frequency and power on the angle $\theta$ (b).}
    \label{fig:frequency}
\end{figure}

The frequency of the parametric radiation generated
by an electron bunch should also be mentioned. Since examined
crystals are two-dimensional, their effective polarizabilities
$g_0$, $g_\tau$ do not depend on the wave vector component that is
parallel to the wires axis \cite{GurnevichNPCS2015}.
In above consideration the angle $\theta$ between the axis $x$
and the bunch velocity direction $\mathbf{v}$ was supposed to be
$90^\circ$ (see Fig.~\ref{fig:frequency}a).
However, it is evident that with the change of this angle
(in case when the ratio $v_y/v_z$ remains constant), the
diffraction geometry will not change. 
At the same time the frequency of parametric radiation will
increase since $k$ grows with the growth of $k_x$.
Calculated dependencies of parametric radiation frequency
and power on the angle $\theta$ are illustrated in the Fig.~\ref{fig:frequency}.
It is obvious that radiation frequency can be
varied in a wide range by the crystal rotation.

Until now we have considered only spontaneous radiation of electron bunches.
However, the induced radiation
can also occur when the electron beam moves through the crystal.
The set of equations describing the interaction of an electromagnetic
wave with the ``crystal-beam'' system, consists of Maxwell's equations 
and those of particle motion in the electromagnetic field.
By analyzing these equations in \cite{Baryshevsky1991} expressions for
generation threshold in case of two-wave diffraction were obtained.
Dependencies of starting (necessary for the induced generation onset)
currents  for crystal built from metallic
wires on the parameter $k_\rho R$ and on the crystal length $L$ (at $k_\rho R\sim 1$)
calculated in accordance with \cite{Baryshevsky1991} are presented below.
The case of Bragg symmetric diffraction was considered. Lets point out that
the starting current for $TE$-polarized wave is substantially lower than for
$TM$-polarized wave. It is related to the fact that in first case $g_0>0$
(Cherenkov radiation is possible), while in second case $g_0<0$. 
\begin{figure}[h]
\begin{minipage}[h]{0.47\linewidth}
\center{\includegraphics[width=0.99\linewidth]{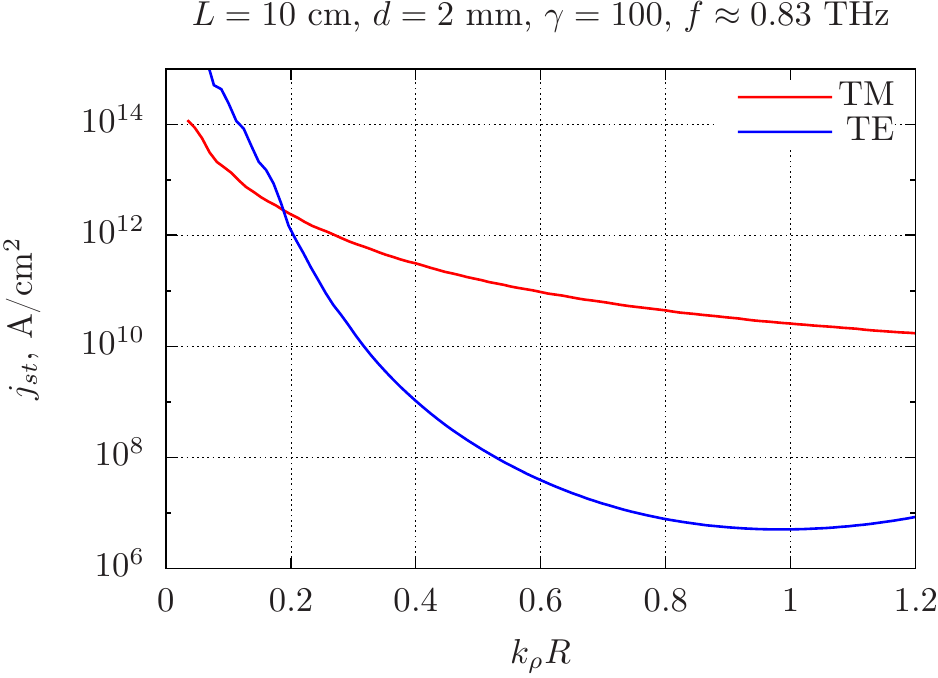}}
\end{minipage}
\hfil
\begin{minipage}[h]{0.47\linewidth}
\center{\includegraphics[width=0.99\linewidth]{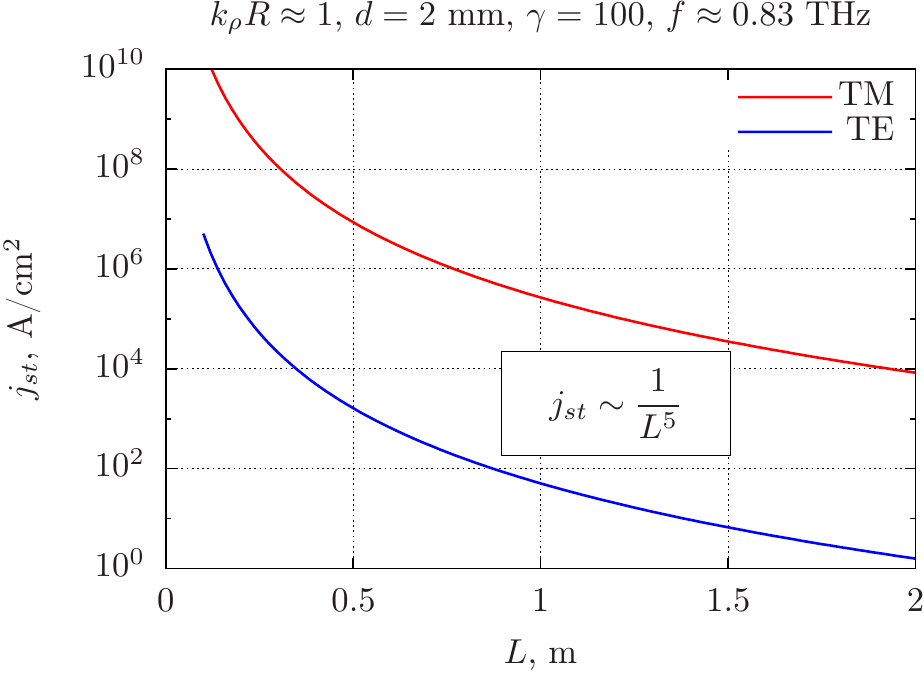}}
\end{minipage}
\label{fig:quasi_cherenkov_power2}
\end{figure}

\section{Conclusion}

We have considered Cherenkov, transition and parametric (quasi-Cherenkov)
radiation emitted by relativistic charged particles passing through a photonic crystal
built from parallel metallic wires.
The radiation emitted in the case when the wavelength becomes
comparable with the wire radius and scattering by a single wire
becomes anisotropic has been analyzed on the basis of the
dynamical theory of diffraction in crystals made from
anisotropically scattering centers.
The dispersion equation derived in this paper enables finding the
possible wave types in the crystal and calculating the unknown
effective polarizabilities $g_\tau$ in the general case (at arbitrary
values of scattering amplitudes $A(\varphi)$).
We have also derived the expressions for
spectral-angular distribution of parametric radiation in
considered crystals.

Numerical solutions of the derived equations for the selected
crystal geometry have confirmed the conclusion
\cite{Baryshevsky2015} that  the intensity of radiation increases
as the radius $R$ of wires is increased, attaining the maximum in
the range $k_\rho R \sim 1$. It has been shown that a considerable
contribution to the total intensity comes from a TE-polarized
wave, but in the case when $k_\rho R \ll 1$, the radiation appears
almost completely TM-polarized. The estimations made here show
that at typical parameters of  modern acceleration facilities, the
radiation intensity attains rather high values, and so
parametric radiation in a crystal built from metallic wires
can be of interest for diverse practical applications, including
the development of high-power THz pulse sources.


\begin{thebibliography}{10}

\bibitem{PXR}
V.~Baryshevsky, I.~Feranchuk, and A.~Ulyanenkov, {\em Parametric X-Ray
  Radiation in Crystals: Theory, Experiment and Applications}.
\newblock Springer, 2005.

\bibitem{Baryshevsky}
V.~Baryshevsky and A.~Gurinovich, ``Spontaneous and induced parametric and
  smith-purcell radiation from electrons moving in a photonic crystal built
  from the metallic threads,'' {\em Nuclear Inst. and Meth. B}, vol.~252,
  no.~1, pp.~92 -- 101, 2006.

\bibitem{Gurnevich2009}
V.~Baryshevsky and E.~Gurnevich, ``The possibility of cherenkov radiation
  generation in a photonic crystal formed by parallel metallic threads,'' {\em
  Vestnik BSU (The Journal of the Belarusian State University), ser.~1, No.3},
  no.~3, pp.~38--44, 2009.

\bibitem{Gurnevich2010}
V.~Baryshevsky and E.~Gurnevich, ``The possibility of cherenkov radiation
  generation in a photonic crystal formed by parallel metallic threads,'' {\em
  Proc. Of 2010 Intern. Kharkov Symp. on Phys. and Engineering of Microwaves,
  Milimeter and Submilimeter Waves (MSMW10), 21-26 June 2010}, pp.~1 -- 3,
  2010.

\bibitem{Baryshevsky2015}
V.~Baryshevsky and A.~Gurinovich, ``Quasi-cherenkov parametric radiation from
  relativistic particles passing through a photonic crystal,'' {\em Nucl. Inst.
  and Meth. B}, vol.~355, pp.~69 -- 75, 2015.
\newblock LANL e-print arXiv:1406.2126.

\bibitem{Vorobev2012}
V.~Vorobev and A.~Tyukhtin, ``Nondivergent cherenkov radiation in a wire
  metamaterial,'' {\em Phys. Rev. Lett.}, vol.~108, p.~184801, 2012.

\bibitem{shiffler2010cerenkov}
D.~Shiffler, J.~Luginsland, D.~French, and J.~Watrous, ``A cerenkov-like maser
  based on a metamaterial structure,'' {\em IEEE Trans. on Plasma Sci.},
  vol.~38, no.~6, pp.~1462 -- 1465, 2010.

\bibitem{smith_purcell}
S.~J. Smith and E.~M. Purcell, ``Visible light from localized surface charges
  moving across a grating,'' {\em Phys. Rev.}, vol.~92, pp.~1069--1069, 1953.

\bibitem{Bolotovsky}
B.~Bolotovskii and G.~Voskresenskii, ``Diffraction radiation,'' {\em Sov. Phys.
  Usp.}, vol.~9, no.~1, p.~73, 1966.

\bibitem{Ter-mikaelyan1960}
M.~Ter-Mikaelian, ``Emission of fast particles in a heterogeneous medium,''
  {\em Doklady Akademii Nauk SSSR}, vol.~34, p.~318, 1960.

\bibitem{Ter_mikaelyan1961}
M.~Ter-Mikaelyan, ``Emission of fast particles in a heterogeneous medium,''
  {\em Nuclear Physics}, vol.~24, pp.~43--61, jan 1961.

\bibitem{bar_molch2009}
V.~Baryshevsky and P.~Molchanov, ``Volume free electron laser with a" grid"
  photonic crystal in a cylindrical waveguide,'' {\em Acta Physica
  Polonica-Series A General Physics}, vol.~115, no.~6, p.~971, 2009.

\bibitem{FEL2009grid}
V.~Baryshevsky, N.~Belous, A.~Gurinovich, E.~Gurnevich, V.~Evdokimov, and
  P.~Molchanov, ``Volume free electron laser with a "grid" photonic crystal
  with variable period: Theory and experiment,'' in {\em Proceedings of
  FEL2009, Liverpool, UK}, 2009.

\bibitem{FirstGridExp1}
V.~Baryshevsky, N.~Belous, A.~Gurinovich, A.~Lobko, P.~Molchanov, and
  V.~Stolyarsky, ``Experimental study of a volume free electron laser with a
  ``grid'' resonator,'' {\em Proc. of FEL 2006 BESSY, Berlin, Germany}, pp.~331
  -- 334, 2006.

\bibitem{radiationRev3}
A.~Tyukhtin and V.~Vorobev, ``Cherenkov radiation in a metamaterial comprised
  of coated wires,'' {\em Journal of the Optical Society of America B},
  vol.~30, p.~1524, may 2013.

\bibitem{radiationRev4}
A.~Tyukhtin and V.~Vorobev, ``Radiation of charges moving along the boundary of
  a wire metamaterial,'' {\em Physical Review E}, vol.~89, jan 2014.

\bibitem{radiationRev6}
V.~Vorobev and A.~Tyukhtin, ``Radiation of a charge moving in wire metamaterial
  perpendicularly to the main axis,'' {\em J. Phys.: Conf. Ser.}, vol.~357,
  p.~012006, may 2012.

\bibitem{radiationRev7}
A.~Tyukhtin, V.~Vorobev, and S.~Galyamin, ``Radiation excited by a
  charged-particle bunch on a planar periodic wire structure,'' {\em Phys. Rev.
  {ST} Accel. Beams}, vol.~17, dec 2014.

\bibitem{BelovReview2012}
C.~Simovski, P.~Belov, A.~Atrashchenko, and Y.~Kivshar, ``Wire metamaterials:
  Physics and applications,'' {\em Advanced Materials}, vol.~24, no.~31,
  pp.~4229 -- 4248, 2012.

\bibitem{Gurnevich2012}
V.~Baryshevsky and E.~Gurnevich, ``Dynamical diffraction theory of waves in
  photonic crystals built from anisotropically scattering elements,'' {\em
  Journal of Nanophotonics}, vol.~6, no.~1, p.~061713, 2012.

\bibitem{baryshevsky2012high}
V.~G. Baryshevsky, {\em High-energy nuclear optics of polarized particles}.
\newblock World Scientific, 2012.

\bibitem{Baryshevsky1997}
V.~Baryshevsky, ``Parametric x-ray radiation at a small angle near the velocity
  direction of the relativistic particle,'' {\em Nucl. Inst. and Meth. B},
  vol.~122, no.~1, pp.~13 -- 18, 1997.

\bibitem{Mors1953}
P.~M. Morse and H.~Feshbach, {\em Methods of Theoretical Physics}.
\newblock Mc Graw Hill, New York, 1953.

\bibitem{Nikolsky}
V.~Nikolskiy and T.~Nikolskaya, {\em Electrodynamics and Radio Waves
  Propagation [in Russian]}.
\newblock Nauka, Moscow, 1989.

\bibitem{BelRev25}
M.~Silveirinha, ``Nonlocal homogenization model for a periodic array of
  $\varepsilon$-negative rods,'' {\em Phys. Rev. E}, vol.~73, p.~046612, Apr
  2006.

\bibitem{Wait1955}
J.~Wait, ``Scattering of a plane wave from a circular dielectric cylinder at
  oblique incidence,'' {\em Canadian Journal of Physics}, vol.~33, no.~5,
  pp.~189--195, 1955.

\bibitem{Janke}
E.~Jahnke, F.~Emde, and F.~L{\"o}sch, {\em Tables of Higher Functions}.
\newblock B.G.~Teubner, Stuttgart, 1966.

\bibitem{landau77quant}
L.~Landau and E.~Lifshitz, {\em Quantum Mechanics: Non-Relativistic Theory}.
\newblock Pergamon Press, 1977.

\bibitem{Pinsker}
Z.~Pinsker, {\em Dynamical Scattering of X-Rays in Crystals}.
\newblock Springer-Verlag, Berlin, 1978.

\bibitem{Ewald1962}
P.~P. Ewald, ed., {\em Fifty Years of X-Ray Diffraction}.
\newblock Springer Science + Business Media, Utrecht, 1962.

\bibitem{James1948}
R.~James, {\em The Optical Principles of the Diffraction of X-Rays}.
\newblock G. Bell \& Sons, London, 1948.

\bibitem{Gurnevich2013arxiv}
V.~Baryshevsky and E.~Gurnevich, ``Multiple scattering of waves in 3d crystals
  (natural or photonic) formed by anisotropically scattering centers,'' {\em
  LANL preprint arXiv:1307.1544}, 2013.

\bibitem{Belov1}
P.~A. Belov, S.~A. Tretyakov, and A.~J. Viitanen, ``Dispersion and reflection
  properties of artificial media formed by regular lattices of ideally
  conducting wires,'' {\em J. of Electromagn. Waves and Appl.}, vol.~16, no.~8,
  pp.~1153--1170, 2002.

\bibitem{Agranovich1984}
V.~M. Agranovich and V.~Ginzburg, {\em Crystal Optics with Spatial Dispersion,
  and Excitons}.
\newblock Springer Berlin Heidelberg, 1984.

\bibitem{bunch_m1}
O.~Lundh, J.~Lim, C.~Rechatin, L.~Ammoura, A.~Ben-Ismail, X.~Davoine,
  G.~Gallot, J.~Goddet, E.~Lefebvre, V.~Malka, {\em et~al.}, ``Few femtosecond,
  few kiloampere electron bunch produced by a laser-plasma accelerator,'' {\em
  Nature Physics}, vol.~7, no.~3, pp.~219 -- 222, 2011.

\bibitem{bunch_m2}
Y.~Liu, X.~Wang, D.~Cline, M.~Babzien, J.~Fang, J.~Gallardo, K.~Kusche,
  I.~Pogorelsky, J.~Skaritka, and A.~Van~Steenbergen, ``Experimental
  observation of femtosecond electron beam microbunching by inverse
  free-electron-laser acceleration,'' {\em Phys. Rev. Lett.}, vol.~80, no.~20,
  p.~4418, 1998.

\bibitem{bunch3}
K.~Kan, J.~Yang, T.~Kondoh, K.~Norizawa, A.~Ogata, T.~Kozawa, and Y.~Yoshida,
  ``Femtosecond electron bunch generation using photocathode rf gun,'' {\em
  Proc. of the FEL2010 Conf., Malmo, Sweden}, pp.~366 -- 369, 2010.

\bibitem{Baryshevsky2011}
V.~Baryshevsky, ``Spontaneous and induced radiation by relativistic particles
  in natural and photonic crystals. crystal x-ray lasers and volume free
  electron lasers (vfel),'' {\em arXiv preprint arXiv:1101.0783}, 2011.

\bibitem{Baryshevsky2015s}
V.~G. Baryshevsky, ``Spontaneous and induced radiation by electrons/positrons
  in natural and photonic crystals. volume free electron lasers ({VFELs}): From
  microwave and optical to x-ray range,'' {\em Nucl. Instrum. Methods Phys.
  Res., Sect. B}, vol.~355, pp.~17--23, jul 2015.

\bibitem{ganter2010swissfel}
R.~Ganter, ``Swissfel-conceptual design report,'' tech. rep., Paul Scherrer
  Institute (PSI), 2010.

\bibitem{BarJPhysFrance1983}
V.~G. Baryshevsky and I.~D. Feranchuk, ``Parametric x-rays from
  ultrarelativistic electrons in a crystal: theory and possibilities of
  practical utilization,'' {\em J. Phys. France}, vol.~44, no.~8, pp.~913--922,
  1983.

\bibitem{GurnevichNPCS2015}
V.~Baryshevsky and E.~Gurnevich, ``Quasi-cherenkov radiation in a photonic
  crystal built from parallel metallic wires in the case of anisotropic
  scattering of waves by the wire,'' in {\em Nonlinear Dynamics and
  Applications (Proc. of NPCS-2015 Conf., Minsk, Belarus)}, vol.~21,
  pp.~126--138, 2015.

\bibitem{Baryshevsky1991}
V.~Baryshevsky, K.~Batrakov, and I.~Dubovskaya, ``Parametric(quasi-cerenkov)
  x-ray free electron lasers,'' {\em Journal of Physics D. Applied Physics},
  vol.~24, pp.~1250--7, 1991.

\end{thebibliography}

\newpage
\appendix
\section{Dispersion equation}\label{appendix_A}

The dispersion equation for a crystal made of parallel metallic
wires that is valid for  $0<k_\rho R \lesssim 1$, has the form
\begin{equation}
\det{D}=0,
\end{equation}
where the elements of matrix   $D$ are determined by the
equalities
\begin{equation}
\begin{aligned}
D_{11} & = (\alpha B_0 + C_1)(iB'_1 + C_7) + i(2\alpha C_9^2-\alpha C_8^2 + \beta C_8C_2) + (3\alpha C_8C_9 - \beta C_9C_2),\\
D_{12} & = -C_3(iB'_1 + C_7) - i\beta C_4 C_8 +\beta C_4 C_9,\\
D_{13} & = C_2(iB'_1 + C_7) + \frac{i}{\beta}C_8(C_1 - (1-\beta^2)C_5) -\frac{1}{\beta}C_9(C_1-(1-\beta^2)C_5) + \frac{\alpha}{\beta}B'_1C_9 - \frac{\alpha}{\beta}C_8C_7,\\
D_{21} & = -C_3,\\
D_{22} & = \frac{\alpha}{1-\beta^2} B_1 + C_5,\\
D_{23} & = -C_4,\\
D_{31} & = \beta^2C_2(iB_0+C_6) + i\beta C_1 C_8 - \beta C_1C_9 - \alpha\beta C_9B_0 - \alpha\beta C_8C_6,\\
D_{32} & = -\beta^2 C_4(iB_0+C_6) - i\beta C_3 C_8 + \beta C_3 C_9,\\
D_{33} & = (\alpha B'_1 + C_1 - C_5 +\beta^2 C_5)(iB_0+C_6) - i(2\alpha
C_9^2-\alpha C_8^2 - \beta C_8C_2) - (3\alpha C_8C_9 + \beta
C_9C_2),
\end{aligned}
\label{eq:DU_coeffs}
\end{equation}
$\alpha=\frac{k_{z0}b}{2\pi}$,  and $\beta=\frac{k_y}{k_\rho}$;
$B_0$, $B_1$ and  $B'_1$ are expressed in terms of the amplitude
of scattering by a wire as
\begin{equation}
\begin{aligned}
B_0 & = \frac{1 + i\pi A_0 - S'_1 A_0}{A_0},\\
B_1 & = \frac{1 + i\pi A_1/2 -S'_3 A_1}{A_1},\\
B'_1 & = \frac{1 + i\pi A_1/2 +(S'_3-S'_1) A_1}{A_1}.\\
\end{aligned}
\end{equation}
The sums  $S'_n$ and  $C_n$ equal
\begin{equation}
\begin{aligned}
S'_1 & = 2\left(\log \frac{k_\rho b}{4\pi} + C\right) -
\sum\limits_{n\in \mathbb{N}_1}\frac{1}{|n|}
+ \sum\limits_{n\in \mathbb{N}_2}\left(\frac{2\pi}{\kappa_{n}b} - \frac{1}{|n|}\right),\\
S'_3 & = \log \frac{k_\rho b}{4\pi} + C -\frac{1}{2} + 
\frac{2\pi^2 + 3k_y^2b^2}{3k_\rho^2b^2} + \frac{2\pi}{k_\rho
b}\sum\limits_{n\in \mathbb{N}_1}\left(\frac{|k_{yn}|}{k_\rho} -
\frac{k_\rho b}{4\pi|n|} \right) + \frac{2\pi}{k_\rho
b}\sum\limits_{n\in \mathbb{N}_2} \left(\frac{|k_{yn}|}{k_\rho} -
\frac{\kappa_{n}}{k_\rho} - \frac{k_\rho b}{4\pi|n|}\right),
\end{aligned}
\end{equation}
\begin{equation}
\begin{aligned}
C_1 & = -\frac{\sin k_{z}a}{\cos k_{z}a - \cos q_za} -
\sum\limits_{n\in \mathbb{N}_1}\frac{\sin k_{zn}a}{\cos k_{zn}a -
\cos q_za}\frac{k_z}{k_{zn}}
-\sum\limits_{n\in \mathbb{N}_2}\left(-1 + \frac{\sinh \kappa_{n}a}{\cosh \kappa_{n}a - \cos q_za}\right)\frac{k_z}{\kappa_{n}},\\
C_2 & = -\frac{\sin k_{z}a}{\cos k_{z}a - \cos q_za} -
\sum\limits_{n\in \mathbb{N}_1}\frac{\sin k_{zn}a}{\cos k_{zn}a -
\cos q_za}\frac{k_zk_{yn}}{k_{zn}k_y}
-\sum\limits_{n\in \mathbb{N}_2}\left(-1 + \frac{\sinh \kappa_{n}a}{\cosh \kappa_{n}a - \cos q_za}\right)\frac{k_zk_{yn}}{\kappa_{n}k_y},\\
C_3 & = \frac{\sin q_{z}a}{\cos k_{z}a - \cos q_za} +
\sum\limits_{n\in \mathbb{N}_1}\frac{\sin q_{z}a}{\cos k_{zn}a -
\cos q_za}
+\sum\limits_{n\in \mathbb{N}_2}\frac{\sin q_za}{\cosh \kappa_{n}a - \cos q_za},\\
C_4 & = \frac{\sin q_{z}a}{\cos k_{z}a - \cos q_za} +
\sum\limits_{n\in \mathbb{N}_1}\frac{\sin q_{z}a}{\cos k_{zn}a -
\cos q_za}\frac{k_{yn}}{k_y}
+\sum\limits_{n\in \mathbb{N}_2}\frac{\sin q_za}{\cosh \kappa_{n}a - \cos q_za}\frac{k_{yn}}{k_y},\\
C_5 & = -\frac{\sin k_{z}a}{\cos k_{z}a - \cos q_za} -
\sum\limits_{n\in \mathbb{N}_1}\frac{\sin k_{zn}a}{\cos k_{zn}a -
\cos q_za}\frac{k_{zn}}{k_{z}} +\sum\limits_{n\in
\mathbb{N}_2}\left(-1 + \frac{\sinh \kappa_{n}a}{\cosh \kappa_{n}a
- \cos q_za}\right)\frac{\kappa_{n}}{k_z},\\
C_6 & = \frac{2\pi}{k_zb}\left( 1+ \sum\limits_{n\in \mathbb{N}_1}
\frac{k_z}{k_{zn}}\right),\;\;\;
C_7 = \frac{2\pi}{k_z b}\frac{k_y^2}{k_\rho^2}\left( 1+ \sum\limits_{n\in \mathbb{N}_1} \frac{k_{z}}{k_{zn}}\frac{k_{yn}^2}{k_y^2}\right),\\
C_8 & = -2\frac{k_y}{k_\rho} + \frac{2\pi}{k_\rho
b}\sum\limits_{n\in \mathbb{N}_2} \frac{k_{yn}}{\kappa_n},\;\;\;
C_9 = \frac{2\pi}{k_z b}\frac{k_y}{k_\rho}\left( 1+ \sum\limits_{n\in \mathbb{N}_1} \frac{k_{z}}{k_{zn}}\frac{k_{yn}}{k_y}\right),\\
\end{aligned}
\label{eq:SC_sums_full}
\end{equation}
where  $\kappa_n=\sqrt{k_{yn}^2-k_\rho^2}$, and summation is
performed over the ranges  $\mathbb{N}_1:\; N_1 \leq n \leq N_2,
\; n\neq 0$, $\mathbb{N}_2:\; n\in (-\infty;
N_1)\cup(N_2,+\infty)$,
$N_1=\left[\frac{(k_y-k_\rho)b}{2\pi}\right]$, and
$N_2=\left[\frac{(k_y+k_\rho)b}{2\pi}\right]$.
Let us note that all sums appearing in \eqref{eq:SC_sums_full} are
real and their values are independent of the
characteristics of the crystal-forming scattering elements
(wires). The functions $S'_1$, $S'_2$, and  $C_6$--$C_9$ are
dependent only on the frequency and the direction of wave
propagation (on $k_y$ and $k_\rho$) and on the crystal periods;
besides that, the functions $C_1$--$C_5$ also depend on
the $z$-component of the wave vector in the
crystal $q_z$. If scattering by a single wire is elastic (which occurs
for  perfectly conducting wires), then $B_0$, $B_1$, and  $B'_1$
are purely real, which can be demonstrated using the optical
theorem, see \cite{Gurnevich2012, landau77quant}. It can be shown
that in this case the solutions $q_z$ of the dispersion equation
are also purely real  (in the transmission band), i.e., no
attenuation will occur for a wave in the crystal. If the wires
have a finite conductivity, the solutions $q_z$ in the general
case will be complex quantities.

Let us consider some cases when the dispersion simplifies
appreciably. Let $k_yb=\pi m$, where $m\neq 0$ is the integer,
i.e., the diffraction conditions in the crystal are fulfilled
exactly for wave vectors $\mathbf{k}$, $\mathbf{k}+\pmb{\tau}$,
where reciprocal lattice vector
$\pmb{\tau}=\pmb{\tau}_y=-\frac{2\pi m}{b}\mathbf{e}_y$. In this
case, the sum $S_2=C_8+iC_9$ (see \eqref{eq:S2_sum}) is
identically equal to zero, and hence the coefficients   $C_8$ and
$C_9$ are also equal to zero. Moreover we can see that
$C_2=C_4=0$. Under such conditions, the dispersion equation can be
written in terms of real variables as follows:
\begin{equation}
(\alpha B'_1 + C_1 -(1-\beta^2)C_5)\left( (\alpha B_0 + C_1)(\alpha B_1 +
(1-\beta^2)C_5) - (1-\beta^2)C_3^2 \right)=0. \label{eq:DU_diffr}
\end{equation}
The approximate analytical solution of this equation for the
simplest cases is found readily. Let, for example,  $k_x=0$,
$k_y=\pi/b$, $ka,kb < 2\pi$, and $\varepsilon_1=\mu_1=1$, whereas
for wave vectors $k$ and $q$  the condition $|q^2/k^2-1|\ll 1$
holds true.
Then we have
\begin{equation}
C_1\approx C_5 \approx -C_3 \approx
-\frac{4}{ka(q^2/k^2-1)}\frac{k_z}{k}, \label{eq:C_simple2}
\end{equation}
and after simple transformations the solutions of
\eqref{eq:DU_diffr} can be presented in the form
\begin{equation}
\left\{
\begin{aligned}
\frac{q_1^2}{k^2} & = 1 + \frac{8\pi}{k^2\Omega_2}
\left\{ \frac{A_0}{1 + i\pi A_0 -S'_1A_0} + \frac{A_1}{1+i\pi A_1/2 - S'_3A_1}\frac{k_z^2}{k^2}\right\},\\
\frac{q_2^2}{k^2} & = 1 +
\frac{8\pi}{k^2\Omega_2}\frac{A_1}{1+i\pi A_1/2
+(S'_3-S'_1)A_1}\frac{k_y^2}{k^2}
\end{aligned}
\right. \label{eq:DU_roots_diffr}
\end{equation}
On the other hand, in the case of two-wave dynamical diffraction
the waves propagating in a photonic crystal are described by the
following set of equations \cite{Baryshevsky2015}:
\begin{equation}
 \left\{
 \begin{aligned}
 & \left( \frac{\pmb{q}^2}{k^2} -1 - g^s_0\right)\mathbf{E}^s(\mathbf{q}) -
  g^s_{-\pmb{\tau}}\mathbf{E}^s(\mathbf{q}+\pmb{\tau}) = 0,\\
 & -g^s_{\pmb{\tau}}\mathbf{E}^s(\mathbf{q}) +
  \left( \frac{(\mathbf{q}+\pmb{\tau})^2}{k^2} -1 - g^s_0\right)\mathbf{E}^s(\mathbf{q}+\pmb{\tau}) = 0,
 \end{aligned}
 \right.
 \label{eq:dynamical_system}
\end{equation}
where $\pmb{\tau}$ is the reciprocal lattice vector, $\mathbf{q}$
is the wave vector in the crystal, the index  $s$ numbers the two
possible polarization states, $g_{\pmb{\tau}}$ are the
coefficients of expansion of the effective dielectric
susceptibility of the crystal into the  Foureir series in terms of
reciprocal lattice vector:
\begin{equation*}
\varepsilon(\mathbf{r}) - 1 = \sum\limits_{\pmb{\tau}} g_{\pmb{\tau}}
 e^{i\pmb{\tau}\mathbf{r}}.
\end{equation*}
The dispersion equation
that follows from the system \eqref{eq:dynamical_system} has a simple form in the
considered case of symmetric Laue diffraction:
\begin{equation*}
(q^2 - k^2(1+g_0))^2 - k^4g_\tau^2 = 0.
\end{equation*}
It has the roots
\begin{equation*}
\frac{q^2_{1,2}}{k^2} = 1 + g_0 \pm g_\tau.
\end{equation*}
Comparing them with \eqref{eq:DU_roots_diffr}, we can find the
unknown quantities  $g_0$ and  $g_\tau$. If the scattering
amplitude is small, then we are led to the result that
agrees well with the conclusions of  \cite{Gurnevich2012}:
\begin{equation}
\left\{
\begin{aligned}
g_0 & \approx \frac{4\pi}{k^2 \Omega_2}\left\{\frac{A_0}{1 + i\pi A_0} + \frac{A_1}{1+i\pi A_1/2} \right\},\\
g_\tau & \approx \frac{4\pi}{k^2 \Omega_2}\left\{\frac{A_0}{1 +
i\pi A_0} + \frac{A_1}{1+i\pi A_1/2}\cos 2\theta_B \right\},
\end{aligned}
\right.
\end{equation}
where $\cos 2\theta_B=\cos^2\theta_B - \sin^2\theta_B =
\frac{k_z^2}{k^2}-\frac{k_y^2}{k^2}$.

The dispersion equation takes an even simpler form in the case
 of normal incidence of a wave onto the crystal ($k_x=k_y=0$).
As $F'_1$  here is identically zero  (see
\eqref{eq:amplitudes_F}), then following the same lines of
reasoning as in section \ref{sec:diffr_and_refr}, instead of the
set of three equations \eqref{eq:Phi_system} we come to a set of
two equations (two first equations in \eqref{eq:Phi_system}, where
we assume $F'_{1n}=0$). In terms of notation  \eqref{eq:DU_coeffs}, the condition of
vanishing the determinant of the system can be written in the form
\begin{equation}
 D_{11}D_{22}-D_{12}D_{21}=0.
\end{equation}
Taking into account that here the coefficients $C_8=C_9=0$, as
well as in the considered case of diffraction, we can write the
explicit form of the  dispersion equation:
\begin{equation}
\frac{k_1b}{2\pi}=-\frac{C_1A_0}{1+i\pi A_0-S'_1 A_0} -
\frac{C_5A_1}{1+i\pi A_1/2-S'_3 A_1}
+\frac{2\pi}{k_1b}\frac{(C_3^2-C_1C_5)A_0A_1}{(1+i\pi A_0-S'_1
A_0)(1+i\pi A_1/2-S'_3 A_1)}. \label{eq:DU_normal_app}
\end{equation}

\end{document}